\documentclass[dvips,11pt,onecolumn,final,fleqn]{article}

\usepackage{graphicx}
\usepackage{amsmath}
\usepackage{amssymb}
\usepackage{subfigure}
\usepackage{textcomp}
\usepackage[in]{fullpage}

\graphicspath{{../EPS/}}

\newcommand{\vect}[1]{\mathbf{#1}}	

\begin{document}


   \begin{flushleft}
      Journal home page: http://www.sciencedirect.com/science/journal/00207462\\
      \vspace{1em}

      Dynamics of a linear oscillator connected to a small strongly non-linear hysteretic absorber\\
      International Journal of Non-Linear Mechanics, Volume 41, Issue 8, October 2006, Pages 969-978\\
      D. Laxalde, F. Thouverez and J.-J. Sinou
   \end{flushleft}
   \vspace{1em}
\begin{center}
    {\LARGE \textbf{Dynamics of a linear oscillator connected to a small strongly non-linear hysteretic absorber}}

   \vspace{1em}
   D. Laxalde$^{a,b}$, F. Thouverez$^{a}$, J.-J. Sinou$^{a}$
   
   \vspace{0.5em}
   ${}^{(a)}$ Laboratoire de Tribologie et Dynamique des Syst\`emes (UMR CNRS 5513)\\\'Ecole Centrale de Lyon, 36 avenue Guy de Collongue, 69134 Ecully Cedex, France\\
   ${}^{(b)}$ Snecma -- Safran group, 77550 Moissy-Cramayel, France

\end{center}
	\begin{abstract}
		The present investigation deals with the dynamics of a two-degrees-of-freedom system which consists of a main linear oscillator and a strongly nonlinear absorber with small mass.
		The nonlinear oscillator has a softening hysteretic characteristic represented by a Bouc-Wen model.
		The periodic solutions of this system are studied and their calculation is performed through an averaging procedure.
		The study of nonlinear modes and their stability shows, under specific conditions, the existence of localization which is responsible for a passive irreversible energy transfer from the linear oscillator to the nonlinear one. 
		The dissipative effect of the nonlinearity appears to play an important role in the energy transfer phenomenon and some design criteria can be drawn regarding this parameter among others to optimize this energy transfer.
		The free transient response is investigated and it is shown that the energy transfer appears when the energy input is sufficient in accordance with the predictions from the nonlinear modes.
		Finally, the steady-state forced response of the system is investigated.
		When the input of energy is sufficient, the resonant response (close to nonlinear modes) experiences localization of the vibrations in the nonlinear absorber and jump phenomena.
	\end{abstract}
\paragraph{Keywords:} Hysteresis, nonlinear energy sinks, non-linear modes, averaging method



\section{Introduction}
Vibration control of mechanical systems is of permanent interest in the field of engineering and research. Dissipation or absorption of the unwanted and often dangerous vibratory energy can be achieved by various ways, using passive or active devices, dissipative materials, coatings, visco-elastic materials, tuned mass dampers or friction \dots

Recently, the interest for nonlinear absorbers kept growing.
In particular, several works on vibration control through the use of a small passive strongly nonlinear device have been presented.
The concept of {\it energy pumping}, that is a passive irreversible one-way energy transfer from a main (linear) structure to a strongly nonlinear (non-linearizable) attachment, was introduced and developed \cite{Vakakis-AppliedMech2,Vakakis-JSV03,Mikhlin-JSV}.
Results on discrete \cite{Vakakis-JSV03,Gendelman-JSV-2005} and continuous systems \cite{Mikhlin-JSV} were presented and in most of studies, the nonlinear absorber consisted in an essential cubic nonlinearity.
Several methodologies where investigated and developed to study this phenomenon. 
The use of asymptotic techniques to find approximate solutions was addressed in several works; Vakakis~{\it et al.~}~\cite{Vakakis-JSV03} used an averaging method, Gendelman {\it et al.}~\cite{Gendelman-JSV-2005} used a multiple scale method, Mikhlin and Reshetnikova~\cite{Mikhlin-JSV} used an expansion method in combining with a Mathieu equation comparison to investigate the stability of periodic solutions\dots
Moreover Vakakis and Rand~\cite{Vakakis-Rand2004} proposed a method to derive exact solutions for systems with cubic nonlinearities based on the use of elliptic functions.
Experimental results were also presented \cite{Vakakis-NLM-exp,Pernot-Exp} in which the nonlinear device is made of a geometric nonlinearity.
Applications have been proposed in shock isolation or in civil engineering.

A common result to explain this phenomenon is that, under specific conditions, some localization of the vibratory can occur leading to an irreversible passive transfer of the energy from the linear structure to the absorber.
This result was demonstrated using the stability of the periodic solutions; Vakakis~\cite{Vakakis-AppliedMech1,Vakakis-AppliedMech2} provided a theoretical background on the subject studying periodic orbits of the associated Hamiltonian system. Several authors, including Vakakis~{\it et al.}~\cite{Vakakis-JSV03} or Mikhlin and Reshetnikova~\cite{Mikhlin-JSV} used the concept of Nonlinear Normal Modes (NNM) and their stability to explain this result.
The concept of nonlinear normal modes was first introduced by Rosenberg \cite{Rosenberg-Ndof} and has been the subject of many investigations in the past years. 
Several authors \cite{Vakakis-NNM,Shaw-Pierre} demonstrated that the use of NNM in studying the dynamics of nonlinear (and in particular {\it strongly} nonlinear) systems has interesting applications both in free and forced responses.
Beside this concept, the effect of localisation of the vibratory energy and motion confinement due to strong nonlinearity was also addressed in several investigations (see Bendiksen~\cite{Bendiksen-Localization} for a good review on the subject).
A main feature of such phenomena is that they are more energy-dependent than frequency-dependent and a direct consequence is that the energy sinks are efficient in a quite wide range of frequency which contrasts with typical linear tuned dampers.

In this paper, we focus on a two degrees-of-freedom system, involving a nonlinear absorber with an hysteretic characteristic.
Hysteretic nonlinearity requires some particular modelling and the one which is used in this study is the Bouc-Wen differential model \cite{Bouc-Hyst,Wen-Hyst}.
A detailed description of the model is addressed in section~\ref{sect:ModelDescription}.
In section~\ref{sect:AsymptoticAnalysis}, an asymptotic method for the study of periodic responses is used. It consists of a two variables expansion combined with an averaging procedure 
The nonlinear modes and their stability, are first studied in section~\ref{sect:NonlinearModes} and numerical results are discussed to highlight to energy pumping phenomenon.
A parametric study emphasize the importance of the dissipation rate of the non-linear absorber and some design criteria are found.
Results on transient response show a correct prediction from the nonlinear modes.
In section~\ref{sect:ForcedResp}, the behaviour of the system in forced response (harmonic excitation) is then addressed and it is shown that, in accordance with the modal prediction, the system experiences some localization phenomenon along with jump phenomena.

\section{Model description} \label{sect:ModelDescription}
A system of two oscillators linearly connected is considered.
The main oscillator, with mass $M$, represents an approximation to some continuous elastic system; the small one, with mass $m$, is the absorber and is strongly nonlinear. 
The coupling is assumed to be weak.
Also, the system remains technologically realistic since the mass ratio about 2\% for the numerical applications.

The motion of this system is governed by the following system of nonlinear equations:
\begin{subequations}\label{eq:mvtR}
	\begin{equation}
	  \ddot{x}(t) + \lambda_0 \dot{x}(t) + \omega_0^2 x(t) + \epsilon_0\left(x(t) - v(t)\right) = f(t)
	\end{equation}
	\begin{equation}
	  \ddot{v}(t) + \lambda_1 \dot{v}(t) + r(\dot v(t),t) + \epsilon_1 \left(v(t)-x(t)\right) = 0
	\end{equation}
\end{subequations}
where $x$ and $v$ are respectively the displacements of the main mass and of the absorber, $\lambda_0$ and $\lambda_1$ are damping ratios, $\omega_0$ is the natural angular frequency of the linear (uncoupled) oscillator, $f(t)$ represents an external forcing and $\epsilon_0$ and $\epsilon_1$ are coupling ratios such that:
\begin{equation}
	\frac{\epsilon_0}{\epsilon_1} = \frac{m}{M}
\end{equation}

The term $r$ represents a nonlinear and hysteretic restoring force which means that it depends of the history of the non-linear motion.

Various kind of systems experience hysteretic behaviours in dynamics; some systems inelastic or with memory may have a restoring force dependent of the history of the deformation, some other, such as rubber or cable isolator are design to dissipate the vibratory energy in the hysteretic loop.
The Bouc-Wen differential model, originally proposed by Bouc \cite{Bouc-Hyst} and reviewed by Wen \cite{Wen-Hyst}, is one of the most used phenomenological model of hysteresis in mechanics. 
It is also used for system identification of hysteretic systems. 
The hysteretic force $r(t)$ is based on the displacement $v(t)$ time history and is given by the following differential equation :
\begin{equation}
	\dot r (t)= A \dot v(t) - \nu \left( \beta |\dot v(t)| |r(t)|^{n-1} r(t) - \gamma \dot v(t) |r(t)|^n \right)
	\label{eq:BoucWen}
\end{equation}
where $A$, $\nu$, $\beta$, $\gamma$ and $n$ are the loop parameters of the Bouc-Wen model.
A proper choice of these parameter allows to describe a wide range of hysteresis loops, with softening or hardening behaviour, different levels of nonlinearity, with various intermediate states possibilities (smooth or bilinear). In what follows, the parameter $n$ is set to $1$.

\section{Asymptotic analysis} \label{sect:AsymptoticAnalysis}
In this section, we will use an averaging method to derive the periodic solutions of a nonlinear system such as system~(\ref{eq:mvtR}).
Let's consider the following general dynamical problem:
\begin{equation}	\label{eq:mvt}
	\ddot{\vect{z}}(t) + g\left(\dot{\vect{z}},\vect{z},t\right)=\vect{0}
\end{equation}
in which $\vect{z}$ is a displacement vector and the term $g$ includes along with linear (stiffness) terms, any nonlinear term and excitations.

We seek a solution of (\ref{eq:mvt}) in the following form:
\begin{equation} \label{eq:solapprox}
   \vect{z}(t)=\vect{z}(\tau,\eta)=\vect{a}(\eta) \cos(\tau +\vect{\varphi}(\eta))
\end{equation}
where the amplitude $\vect{a}$ and phase $\vect{\varphi}$ are slowly varying quantities (time scale $\eta$) with respect to the fast time scale $\tau$. 
This transformation from the displacement dependent variables (one variable respectively) to the amplitudes and phases dependent variables (two variables respectively) allows us to impose an additional condition. 
It is usual to choose that the velocity has a similar form to the linear case:
\begin{equation} \label{eq:CompDeplVit1}
	\dot{\vect{z}} =\frac{d{\vect{z}}(\tau,\eta)}{dt} = -\frac{d\tau}{dt}{\vect{a}}(\eta) \sin(\tau +\varphi(\eta))
\end{equation}
The time scale $\tau$ can then be interpreted as the time scale of the periodic motion whereas, the time scale $\eta$ represents a perturbation time scale.
Also, in equation~(\ref{eq:CompDeplVit1}), the term:
\begin{equation} \label{eq:Omega}
	\frac{d\tau}{dt}=\omega
\end{equation}
represents an angular-frequency-like "variable" assumed to be constant in time.
However, this angular frequency $\omega$ can be amplitude-dependent and \textit{a priori} unknown as in the case of nonlinear modes (see section~\ref{sect:NonlinearModes}) or not as in the case of steady-state forced response where it corresponds to the excitation frequency (see section~\ref{sect:ForcedResp}).

Equation (\ref{eq:CompDeplVit1}) now becomes:
\begin{equation} \label{eq:Vit}
	\dot{\vect{z}} = -\omega \vect{a}(\eta) \sin(\tau +\vect{\varphi}(\eta))
\end{equation}

Differentiating (\ref{eq:solapprox}) with respect to time and equating the result with (\ref{eq:Vit}), we find:
\begin{equation} \label{eq:Slow1}
	\vect{a}^{\prime} \cos(\tau +\vect{\varphi}) - \vect{a} \vect{\varphi}^{\prime} \sin(\tau +\vect{\varphi}) = \vect{0}
\end{equation}
where ${.}^{\prime}$ denotes derivatives with respect to the slow time scale $\eta$.

We have obtained the first equation governing the evolution of the slow flow variables $\vect{a}$ and $\vect{\varphi}$.
The second one can be obtained by first differentiating (\ref{eq:Vit}) with respect to $t$, which yields:
\begin{equation}	\label{eq:Accel}
	\ddot{\vect{z}} = -\omega \vect{a}^{\prime} \sin(\tau +\vect{\varphi})  - \omega \vect{a} \vect{\varphi}^{\prime} \cos(\tau +\vect{\varphi})  - \omega^2 \vect{a} \cos(\tau +\vect{\varphi})
\end{equation}
and substituting $\ddot{\vect{z}}$ into (\ref{eq:mvt}):
\begin{equation} \label{eq:Slow2}
	-\omega \vect{a}^{\prime} \sin(\tau +\vect{\varphi})  - \omega \vect{a} \vect{\varphi}^{\prime} \cos(\tau +\vect{\varphi})  - \omega^2 \vect{a} \cos(\tau +\vect{\varphi}) + g(\vect{a} \cos (\tau +\vect{\varphi}),-\omega \vect{a} \sin(\tau +\vect{\varphi}))=\vect{0}
\end{equation}

Finally, solving equations (\ref{eq:Slow1}) and (\ref{eq:Slow2}) for slow flow variables variations $\vect{a}^{\prime}$ and $\vect{\varphi}^{\prime}$ we have the following system:
\begin{subequations} \label{eq:SlowStand}
   \begin{equation}
      \vect{a}^{\prime}=\dfrac{\sin(\tau +\vect{\varphi})}{\omega} \left(G(\vect{a},\vect{\varphi},\tau)-\omega^2 \vect{a}\cos(\tau +\vect{\varphi})\right)
   \end{equation}
   \begin{equation}
      \vect{a}\vect{\varphi}^{\prime}=\dfrac{\cos(\tau +\vect{\varphi})}{\omega} \left(G(\vect{a},\vect{\varphi},\tau)-\omega^2 \vect{a}\cos(\tau +\vect{\varphi})\right)
	\end{equation}
\end{subequations}
where $G(\vect{a},\vect{\varphi})$ is obtained by substituting $\vect{z}$ and $\dot{\vect{z}}$ in $g$.

The system (\ref{eq:SlowStand}) is now in standard form and can be averaged over the fast time scales $\tau$ with the slow flow variables $\vect{a}$ and $\vect{\varphi}$ being taken as constants.
\begin{subequations} \label{eq:SlowMoy}
	\begin{equation}
		\vect{a}^{\prime} = \dfrac{1}{\omega} \dfrac{1}{2\pi}\displaystyle \int_0^{2\pi}G(\vect{a},\vect{\varphi},\psi)\sin \psi d\psi
     \end{equation}
     \begin{equation}
		\vect{a}\vect{\varphi}^{\prime} = \dfrac{1}{\omega} \left[ \dfrac{1}{2\pi}\displaystyle \int_0^{2\pi} G(\vect{a},\vect{\varphi},\psi) \cos\psi d\psi-\dfrac{\omega^2}{2} \vect{a}\right]
	\end{equation}
\end{subequations}
The averaged problem defined by the differential system (\ref{eq:SlowMoy}) provides approximate solutions with the particular form defined by (\ref{eq:solapprox}). 
The approximation is made on the form of the nonlinear terms in $g$ which are assumed (due to averaging) to be proportional to the harmonic functions $\cos \psi$ and $\sin \psi$.

This formalism will be used in the following section to derive particular solutions to different problems, including free vibrations or forced vibrations, by substituting the proper function $g$ in equations~(\ref{eq:SlowMoy}).

\vspace{1ex}
In the system subject of this paper, the general term $g(\vect{z},\dot{\vect{z}},t)$ includes, along with linear terms, nonlinear dissipative terms defined by the differential Bouc-Wen model.
As these terms are difficult to average in close form, we have used a numeric approach.
However, in order to proceed with the analytical calculations, we introduce a similar notation to equation~(\ref{eq:solapprox}) for the averaged nonlinear restoring force $r$ of the Bouc-Wen model.
The resulting amplitude $a_r$ and phase $\varphi_r$ are derived numerically by substituting $v(t)=a_v\cos(\tau+\varphi_v)$ in relation~(\ref{eq:BoucWen}) and performing an averaging on the fast time scale (as in equation~(\ref{eq:SlowMoy})):
\begin{equation}    \label{eq:TermNLHarm}
	a_r e^{i\varphi_r} = \dfrac{1}{2\pi} \int_0^{2\pi} r(\dot v,t) e^{-i\psi} d\psi
\end{equation}
This step introduce no additional approximation or simplification to the averaging procedure of equation~(\ref{eq:SlowMoy}).
Details on the variations of the averaged hysteretic restoring force with the non-linear displacement $v$ are provided in section~\ref{ssect:NLM_Num} where the equivalent stiffness and damping are studied.

\section{Non-linear modes} 	\label{sect:NonlinearModes}
In section, we perform a modal analysis of the system~(\ref{eq:mvtR}) unforced and undamped.
Modal studies of non-linear systems are numerous in the literature~\cite{Rosenberg-Ndof,Shaw-Pierre,Vakakis-NNM,Wanda,Wanda-NLM} and the theoretical concept of non-linear modes is found to provide interesting dynamical descriptions as well as a valuable design tool.
The derivation of non-linear modes is first performed analytically.
To do so, we first apply the averaging method described in section~\ref{sect:AsymptoticAnalysis} to the initial equations of motion~(\ref{eq:mvtR}) (without forcing and damping); then, the study of the fixed points of the resulting system leads to the definition of an eigenvalue problem.
In a second time, numerical results are presented for several types of Bouc-Wen hysteretic restoring forces and to possibility of localization phenomenon and possible energy pumping are demonstrated.

\subsection{Free vibrations analysis -- Non-linear modes} 	\label{ssect:FreeVib_NLM}
The free vibrations problem of the two degrees of freedom system is defined by:
\begin{subequations}\label{eq:mvt_free}
   \begin{equation}
      \ddot{x}(t) + \omega_0^2 x(t) + \epsilon_0\left(x(t)- v(t)\right)  = 0
   \end{equation}
   \begin{equation}
      \ddot v(t)  + r(\dot v(t),t)+ \epsilon_1 \left(v(t)-x(t)\right)  = 0 
   \end{equation}
\end{subequations}

Using the averaging method described in section~\ref{sect:AsymptoticAnalysis}, the two displacements variables take the form of (\ref{eq:solapprox}), {\it ie}:
\begin{subequations}
   \begin{equation}
      x=a_x(\eta) \cos(\tau +\varphi_x(\eta))
   \end{equation}
   and
   \begin{equation}
      v=a_v(\eta) \cos(\tau +\varphi_v(\eta))
   \end{equation}
   \label{eq:NLM-solapprox}
\end{subequations}
where $\tau=\omega t$, and $\omega$ is the angular frequency of the oscillations.

The term $G(a,\varphi,\tau)$ in equations~(\ref{eq:SlowStand}) can be simply derived and introduced in equations~(\ref{eq:SlowMoy}), which leads to:
\begin{subequations} \label{eq:SlowMoyFree1}
   \begin{equation}
      2\omega a_x^{\prime} = -\epsilon_0a_v\sin(\varphi_x-\varphi_v)
   \end{equation}
   \begin{equation}
      2\omega a_v^{\prime} = \epsilon_1a_x\sin(\varphi_x-\varphi_v) -\dfrac{1}{m} a_r\sin(\varphi_r-\varphi_v)
   \end{equation}
   \begin{equation}
      2\omega a_x \varphi_x^{\prime} = (\omega_0^2+\epsilon_0-\omega^2)a_x -\epsilon_0 a_v\cos(\varphi_x-\varphi_v)
   \end{equation}
   \begin{equation}
      2\omega a_v \varphi_v^{\prime} = (\epsilon_1-\omega^2)a_v -\epsilon_1a_x\cos(\varphi_x-\varphi_v) +\dfrac{1}{m} a_r\cos(\varphi_r-\varphi_v)
   \end{equation}
\end{subequations}

As the system~(\ref{eq:SlowMoyFree1}) is in standard first-order form, we can study its fixed points by making all time derivatives zero (left hand side).
Then one obtains a system of nonlinear equations (with amplitudes $a_j$ and phases $\varphi_j$ along with the angular frequency $\omega$ as unknowns) which can be interpreted as a non-linear eigenvalue problem to find an approximation of the non-linear modes.

However, from the first equation of system~(\ref{eq:SlowMoyFree1}), one can notice that the fixed points necessary verify, 
\begin{equation}
	\varphi_x-\varphi_v = 0 \mod(\pi)
\end{equation}
Carrying this result in the second equation, one finds that 
\begin{equation}
	a_r\sin(\varphi_v-\varphi_r)=0
\end{equation}
which is absurd because first $a_r$ cannot be zero and second because the hysteretic and dissipative characteristics of the nonlinearity impose a non-zero phase difference between the restoring force $r$ and the displacement $v$.

In order to avoid this, the general form of the solutions (given by equations~(\ref{eq:solapprox}) or (\ref{eq:NLM-solapprox})) needs to be adapted. The formalism was inspired by the complex mode definition for linear systems and is given by:
\begin{equation} \label{eq:FormComplexModes}
	\vect{z}(\tau,\eta)=\vect{c}(\eta) e^{-\zeta \eta} \cos(\tau +\vect{\varphi}(\eta))
\end{equation}
where $\zeta$ represents the modal damping and $\vect{c}$ is the new amplitude variable.
The exponential decay allows a proper modelling of the dissipative  aspect of the motion and can be related with a complex natural frequency (by analogy with linear complex modes),
\begin{equation}
	\lambda = - \zeta \pm i \omega
\end{equation}
It was also assumed that this exponential decay is slowly varying.

Now, introducing the formalism of complex modes, defined by equation~(\ref{eq:FormComplexModes}), in system~(\ref{eq:SlowMoyFree1}) and noting $a_i(\eta)=c_i(\eta) e^{-\zeta \eta}$ for $i=x,v$, one obtains:
\begin{subequations} \label{eq:CNM}
   \begin{equation}
      c_x^{\prime} = \zeta c_x - \dfrac{\epsilon_0}{2\omega}c_v\sin(\varphi_x-\varphi_v)
   \end{equation}
   \begin{equation}
      c_v^{\prime} = \zeta c_v  + \dfrac{\epsilon_1}{2\omega}c_x\sin(\varphi_x-\varphi_v) +\dfrac{1}{2\omega} c_r\sin(\varphi_v-\varphi_r)
   \end{equation}
   \begin{equation}
      c_x \varphi_x^{\prime} = \frac{\omega_0^2+\epsilon_0-\omega^2}{2\omega}c_x -\dfrac{\epsilon_0}{2\omega} c_v\cos(\varphi_x-\varphi_v)
      \label{eq:CNM3}
   \end{equation}
   \begin{equation}
      c_v \varphi_v^{\prime} = \frac{\epsilon_1-\omega^2}{2\omega}c_v -\dfrac{\epsilon_1}{2\omega}c_x\cos(\varphi_x-\varphi_v) +\dfrac{1}{2\omega}c_r\cos(\varphi_v-\varphi_r)
   \label{eq:CNM4}
   \end{equation}
\end{subequations}
It should be noticed that along with the assumptions made on the form of the displacements, equation~(\ref{eq:FormComplexModes}), the nonlinear term $a_r(\eta)$ is assumed to have the same exponential decay that the displacement. This assumption can be justified by looking at the expression of the averaged nonlinear term.
In equation~(\ref{eq:TermNLHarm}), if we substitute for $\dot v$ the new complex form (\ref{eq:FormComplexModes}), the exponential decay is reported outside the summation symbol and as a consequence, equation~(\ref{eq:TermNLHarm}) can be updated with the complex form defined by (\ref{eq:FormComplexModes}).

If we finally combine equations~(\ref{eq:CNM3}) and (\ref{eq:CNM4}) and introduce the phase difference variables $\varphi_{ij}=\varphi_i - \varphi_j$, with $i,j=x,v \mbox{ or } r$, we obtain the following system:
\begin{subequations} \label{eq:CNMr}
   \begin{equation}
      c_x^{\prime} = \zeta c_x - \dfrac{\epsilon_0}{2\omega}c_v \sin \varphi_{xv}
   \end{equation}
   \begin{equation}
      c_v^{\prime} = \zeta c_v  + \dfrac{\epsilon_1}{2\omega}c_x \sin \varphi_{xv} +\dfrac{1}{2\omega} c_r \sin \varphi_{vr}
   \end{equation}
   \begin{equation}
      c_x c_v \varphi_{xv}^{\prime} = \frac{\omega_0^2+\epsilon_0-\epsilon_1}{2\omega}c_x c_v -\dfrac{1}{2\omega}\left(\epsilon_0 c_v^2 -\epsilon_1 c_x^2 \right)\cos \varphi_{xv}-\dfrac{1}{2\omega} c_r \cos \varphi_{vr}
   \end{equation}
\end{subequations}

This unforced system is in (averaged) standard form and its fixed points are the main approximation of the nonlinear modes.

Along with this, we have to introduce an additional relation between the coordinates in order to normalize the modes. To do so, we define $\mathcal{H}$ the global energy of the system by:
\begin{equation}	\label{eq:Energy}
	\mathcal{H} = \mathcal{T} + \mathcal{U} -\mathcal{W}_d
\end{equation}
where $\mathcal{T}$, $\mathcal{U}$ are respectively the global kinetic and potential energies of the system; $\mathcal{W}_d$ is the energy dissipated by the hysteretic force during one cycle of the motion.

Also note that, if the non-linear modes can be approximated by the fixed points of equations~(\ref{eq:CNMr}), these will depend on the value of the angular frequency $\omega$ which is an unknown of the free dynamical problem.
This angular frequency $\omega$, which represents the eigenfrequency of the modes, is determined using the additionnal energy relation~(\ref{eq:Energy}).

This formalism introduces the notion of complex nonlinear modes which correspond (as in the linear domain) to special solutions of the free vibrations problem where the system's degrees of freedom oscillate with the same frequency but, in contrast with non-linear \textit{normal} modes, with a phase difference between them.
The dissipative terms introduce a phase difference between the coordinates and a \textit{modal} damping $\zeta$ which, as the natural frequencies and deformed shapes depends on the system's energy. 
It may also be note that, in contrast with classical nonlinear normal modes, these complex modes as defined here are not necessary normal to the iso-energetic curves.
With this complex modal description, one can handle the dissipative effects of the nonlinearity directly without any assumption regarding their importance.
This particular feature will be used in the following numerical applications to determine design rules on the dissipation rates.
This is a major concern, in engineering applications, when dealing with very weakly damped (linear) structure; in these situations, the nonlinear oscillator has to ensure the resonance capture phenomenon as well as the dissipation of the vibratory energy.

\subsection{Stability}
The stability of the nonlinear modes can be determined using the system~(\ref{eq:CNMr}), which is in standard form, and studying the eigenvalues of the Jacobian matrix $D_{\vect{z}}F(\vect{z_0},M)$ at the equilibrium points.
\begin{multline}    \label{eq:Jacob}
	D_{\vect{z}}F(\vect{z_0},M)=\\
	\begin{bmatrix}
		\zeta & -\dfrac{\epsilon_0}{2\omega}\sin \varphi_{xv} & - \dfrac{\epsilon_0}{2\omega}c_v\cos \varphi_{xv} \\[0.5cm]
		\dfrac{\epsilon_1}{2\omega}\sin \varphi_{xv} & \zeta + \dfrac{1}{2\omega} \dfrac{\partial (c_r\sin \varphi_{vr})}{\partial c_v} & \begin{array}{l} \dfrac{\epsilon_1}{2\omega}c_x\cos \varphi_{xv}\\\hspace{1.2cm}  + \dfrac{1}{2\omega} \dfrac{\partial (c_r\sin\varphi_{vr})}{\partial \varphi_{xv}}\end{array}\\[0.5cm]
			\dfrac{\omega_0^2+\epsilon_0-\epsilon_1}{2\omega} c_v +\dfrac{\epsilon_1}{\omega}c_x\cos \varphi_{xv} &  \begin{array}{l} \dfrac{\omega_0^2+\epsilon_0-\epsilon_1}{2\omega}c_v -\dfrac{\epsilon_0}{\omega}c_v\cos \varphi_{xv}\\\hspace{1.2cm} - \dfrac{1}{2\omega} \dfrac{\partial c_r \cos \varphi_{vr}}{\partial c_v}\end{array} & \begin{array}{l} \dfrac{1}{2\omega}\left(\epsilon_0 c_v^2 -\epsilon_1 c_x^2 \right)\sin \varphi_{xv}\\ \hspace{1cm}-\dfrac{1}{2\omega} \dfrac{\partial(c_r \cos \varphi_{vr})}{\partial \varphi_{xv}}\end{array}
	\end{bmatrix}
\end{multline}

\subsection{Numerical results, localization and energy pumping} 	\label{ssect:NLM_Num}
\begin{figure}[ht]
	\begin{center}
		\subfigure[]{\includegraphics[width=0.48\textwidth]{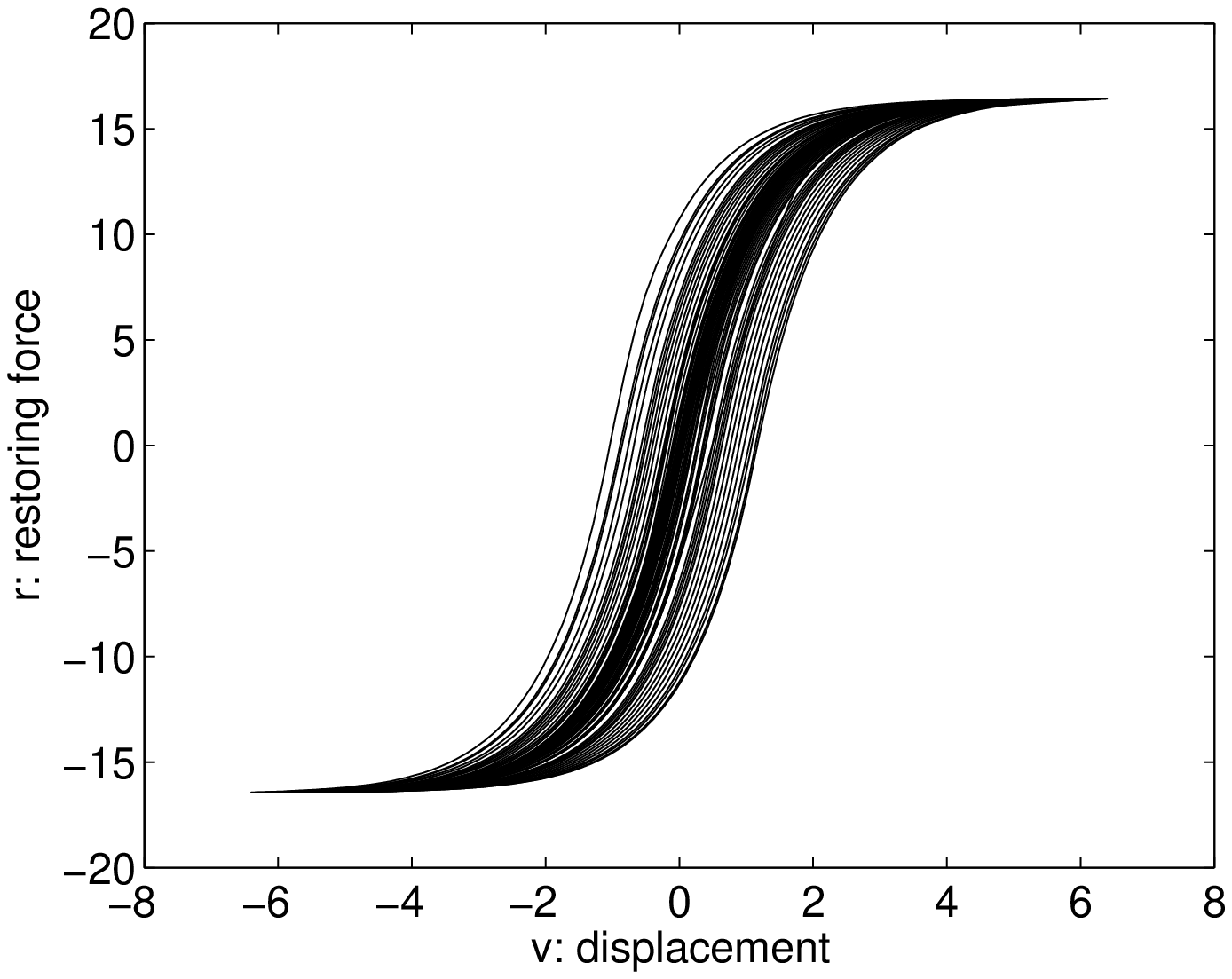}}
		\subfigure[]{\includegraphics[width=0.48\textwidth]{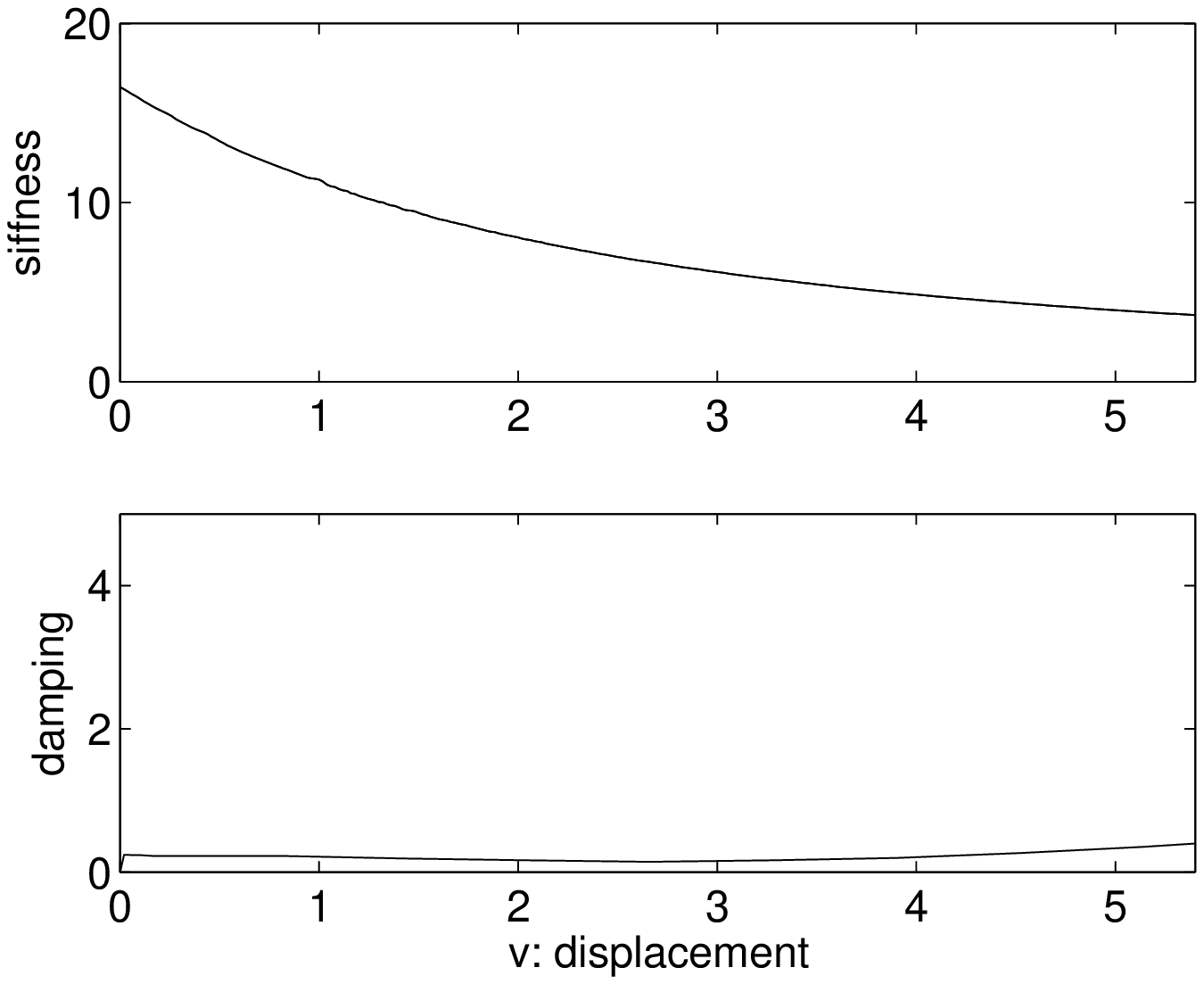}}
	\end{center}
	\caption[Example n\textdegree 1 - $\gamma=10^{-3}$]{Example n\textdegree 1 - $\gamma=10^{-3}$; (a)~hysteretic cycle, (b)~equivalent stiffness and damping.}
	\label{fig:Cycle1}
\end{figure}

\begin{figure}[ht]
	\begin{center}
		\subfigure[]{\includegraphics[width=0.48\textwidth]{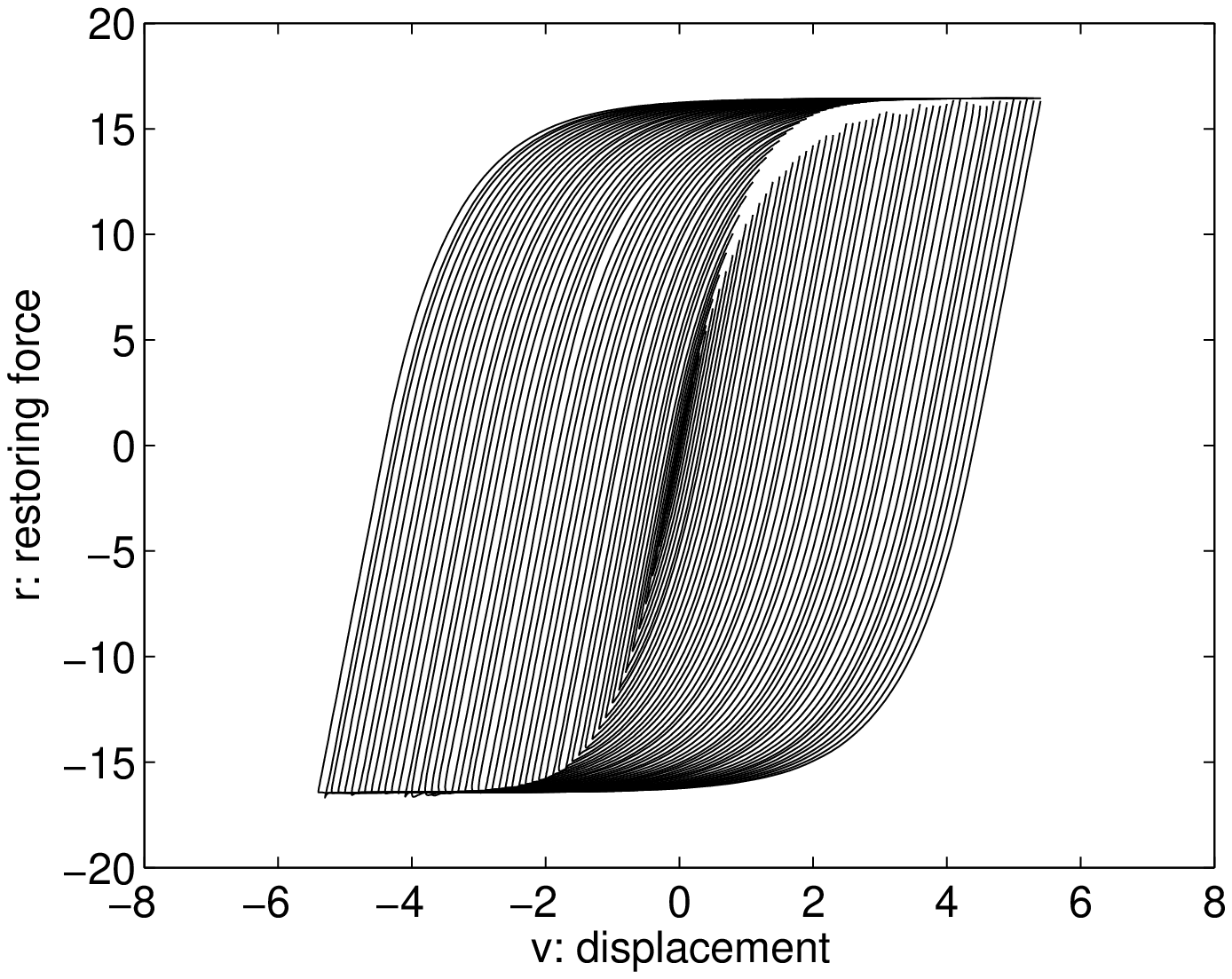}}
		\subfigure[]{\includegraphics[width=0.48\textwidth]{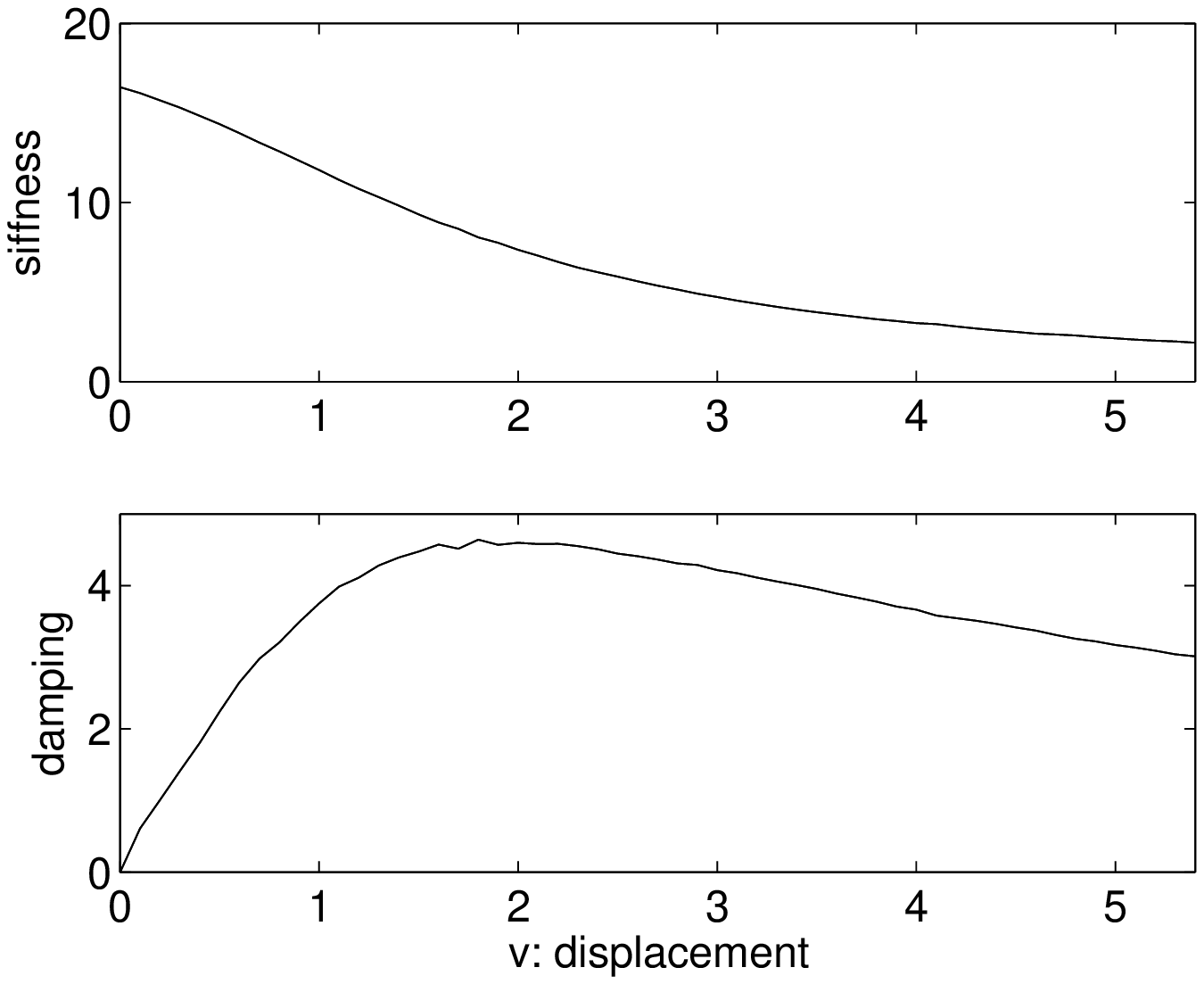}}
	\end{center}
	\caption[Example n\textdegree 2 - $\gamma=0.5$]{Example n\textdegree 2 - $\gamma=0.5$; (a)~hysteretic cycle, (b)~equivalent stiffness and damping.}
	\label{fig:Cycle2}
\end{figure}

The hysteretic Bouc-Wen model is quite versatile and a large variety of hysteretic loops can be modelled.
In this study, we focused on the class of softening hysteretic nonlinearity (\textit{i.e.} $\gamma=-0.5\leqslant0$ for example in equation~(\ref{eq:BoucWen})) and we particularly investigated the influence of the (nonlinear) dissipation due to hysteresis.
As an example, let's consider the two hysteretic loops depicted in figures~\ref{fig:Cycle1} and \ref{fig:Cycle2} and their equivalent stiffness and damping (first harmonic).
Both of these examples feature a strongly nonlinear behaviour.
Their equivalent stiffness are quite similar as opposed to their equivalent damping.
Actually as the loop area (which is directly related to the energy dissipation) of the second example is larger, its equivalent damping is consequently more important.

In this section, some numerical results on the nonlinear modes on these two representative examples are presented.
The nonlinear complex modes are derived by numerically solving the fixed points problem defined by system~(\ref{eq:CNMr}) along with the energy relationship given in equation~(\ref{eq:Energy}) which aims at finding the unknown eigenfrequency $\omega$ and at normalizing the modes.
Since our motivation is to investigated the vibration control of weakly damped structures, the main oscillator is assumed to be conservative and the only source of dissipation is the absorber.

\subsubsection*{Example 1.}
\begin{figure}[htbp]
    \centering
    \subfigure[Linear oscillator]{\includegraphics[width=.49\textwidth]{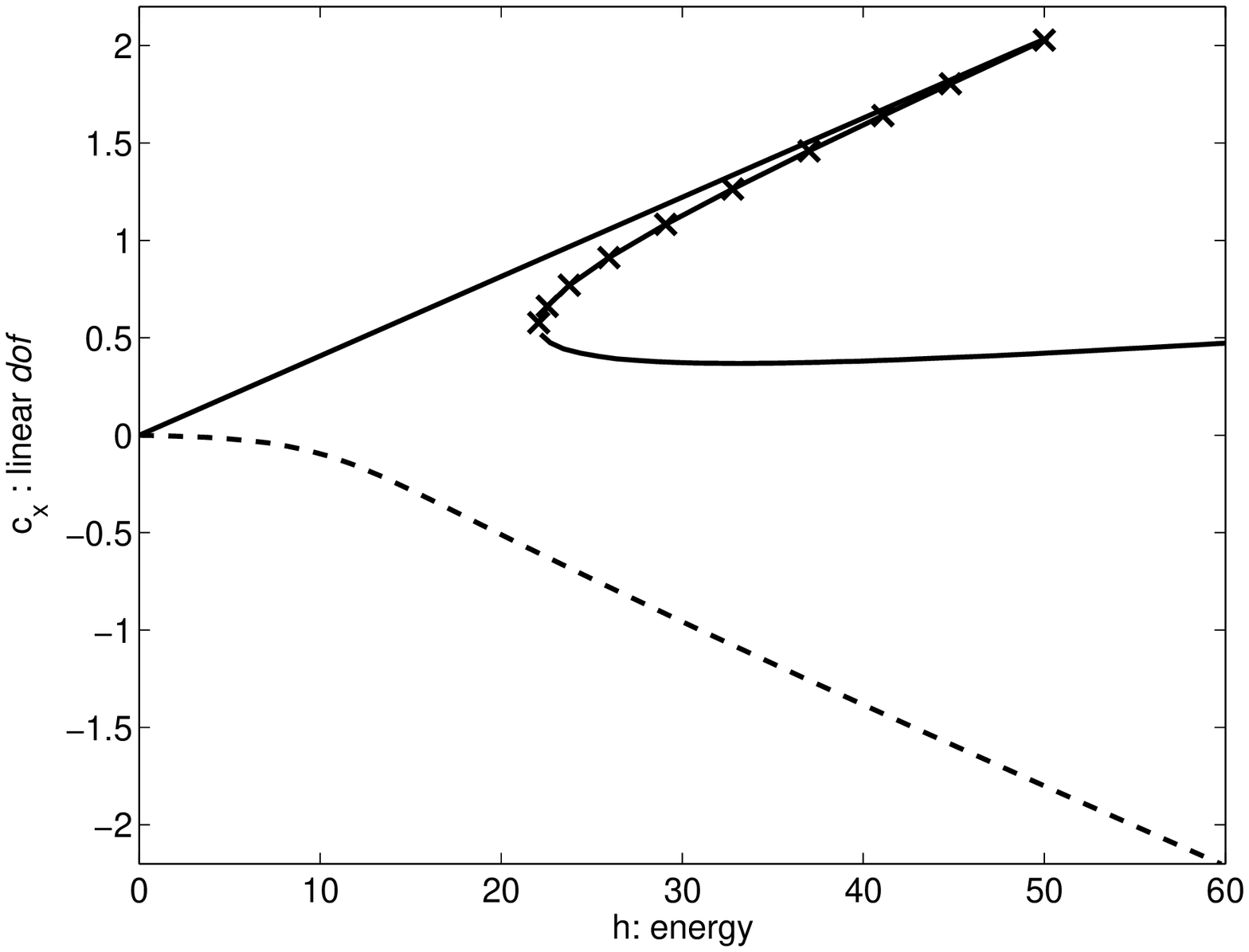}}
    \subfigure[Absorber]{\includegraphics[width=.49\textwidth]{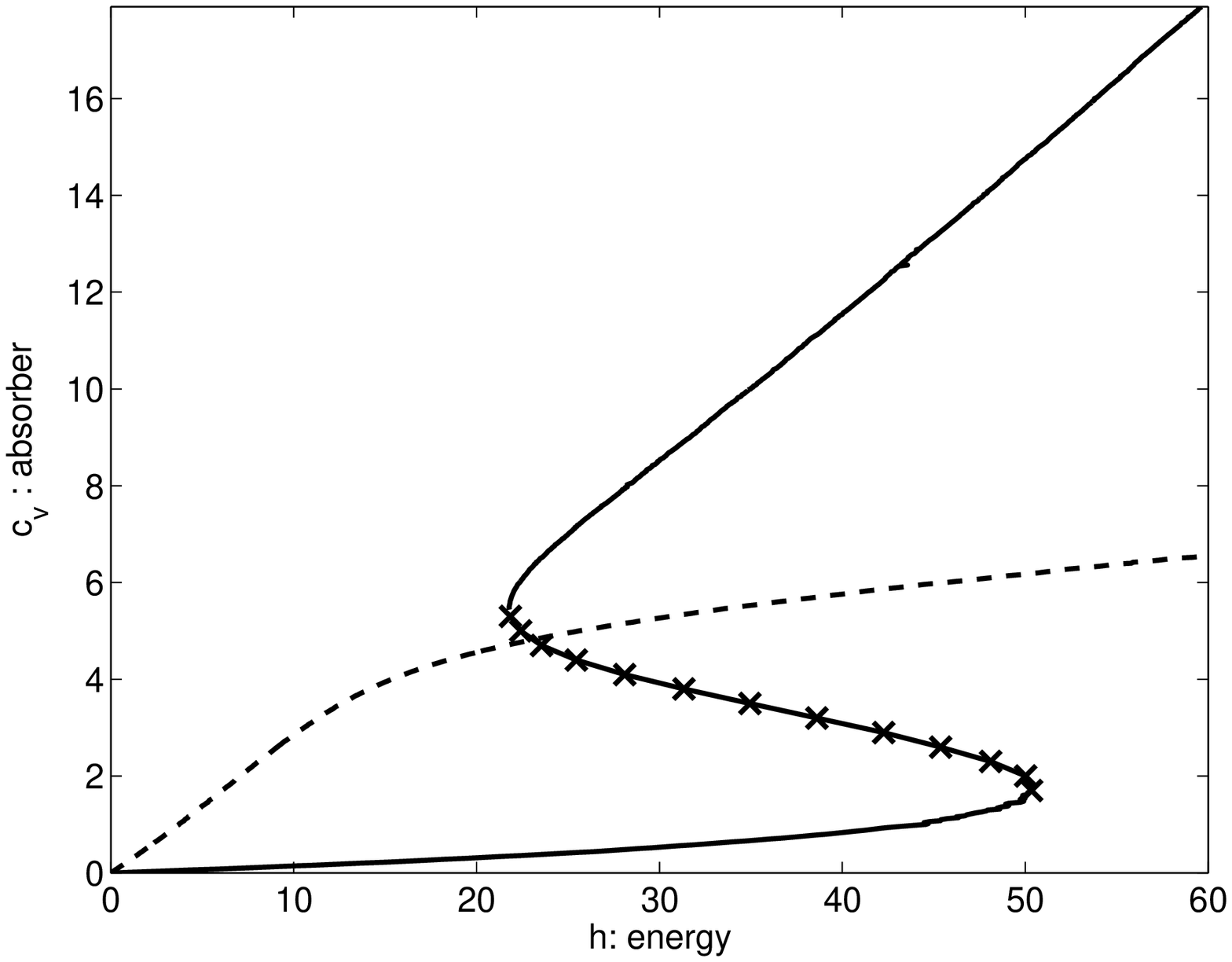}} 
    \subfigure[Eigenfrequency]{\includegraphics[width=.49\textwidth]{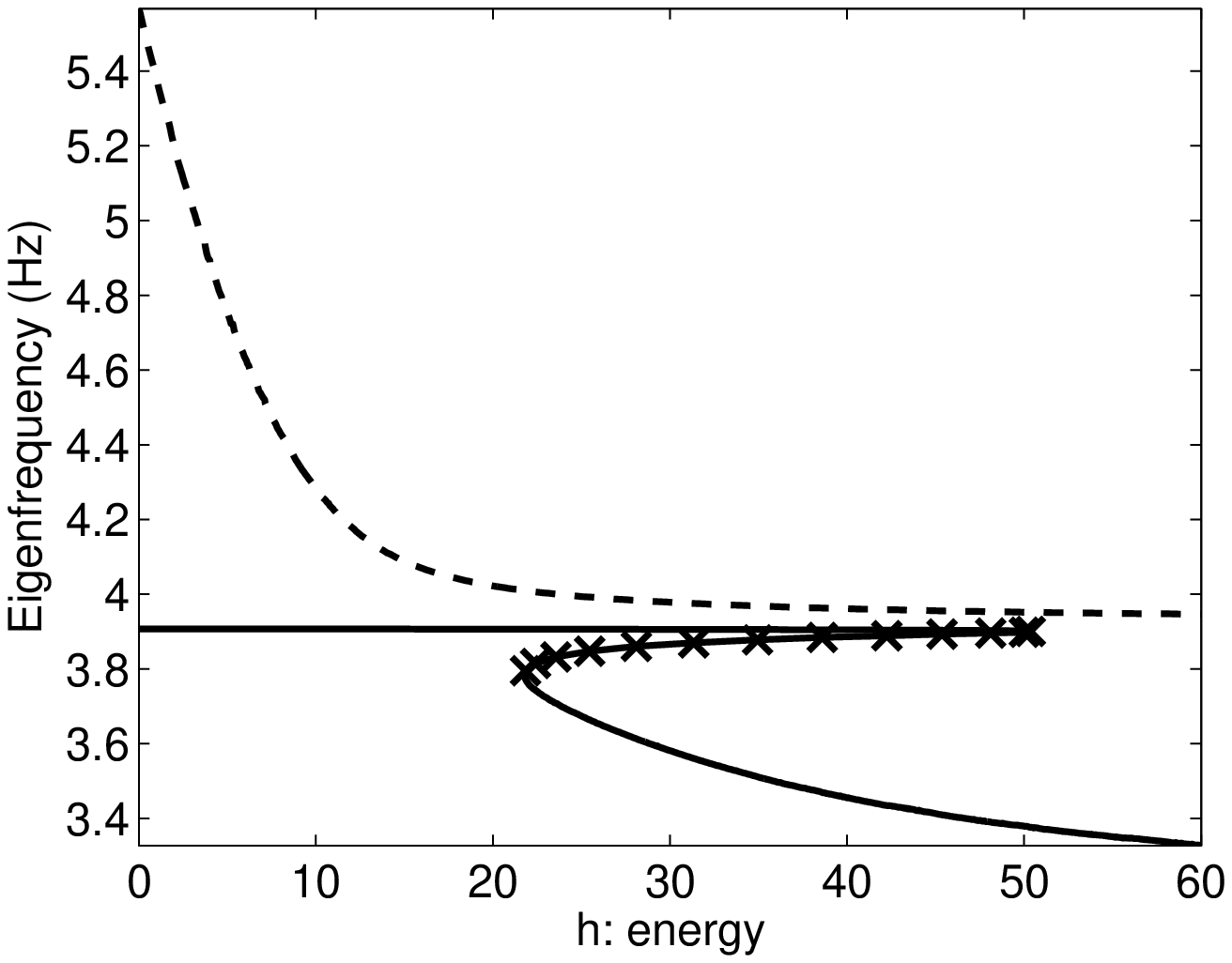}}
    \subfigure[Modal damping]{\includegraphics[width=.49\textwidth]{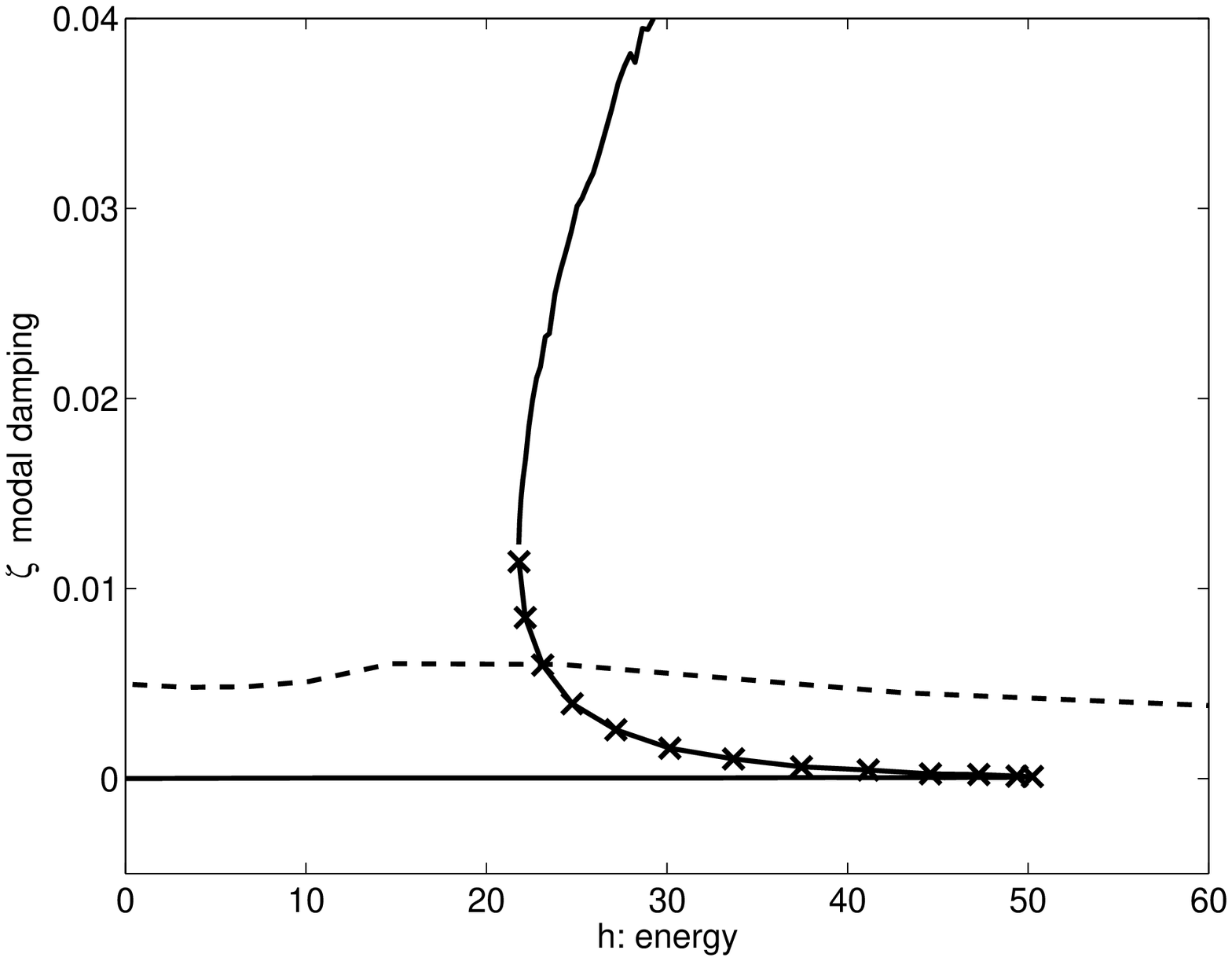}}
    \caption[Nonlinear modes - Example 1]{Example 1: Nonlinear modes: (---, - - -), stable; (-x-x-), unstable.}
    \label{fig:ModesNL1}
\end{figure}

We first study the case of a weakly dissipative hysteretic cycle which build using the Bouc-Wen parameters $\beta=10^{-3}$ and $\gamma=-0.5$.
The modal quantities are depicted in figures~\ref{fig:ModesNL1}.
The modal amplitudes of the linear and nonlinear degrees of freedom, $c_x$ and $c_v$, natural frequency $\omega$, and modal damping $\zeta$ are plotted versus the global energy $\mathcal{H}$ of the system.

There are two main branches (in solid and dashed lines) and an unstable region appears in the solid line branch symbolized by crosses.
We can discriminate several regions with distinct dynamical behaviours: 
\begin{description}
	\item[Low energies region: ] in the solid line branch, the motion appears to be nearly localized in the linear oscillator which modal curve is a straight line: the nonlinear system's behaviour is linearizable.
Note that, in this region, the natural frequency of the linear oscillator is constant.
The dashed line branch shows, on the other hand, a prominent motion in the absorber and a decreasing natural frequency due to the softening characteristic of the nonlinearity.
	\item[High energies region: ] there, the situation is quite reversed, the solid line branch shows a strong localization in the nonlinear oscillator with a decreasing natural frequency; and the dashed line branch displays a prominent motion in the linear system.
	\item[Intermediate energy region: ] between the previous asymptotic states, the system experiences a bifurcation phenomenon of its modes.
		When the global energy increases, the motion of the linear oscillator jumps from a high to a low level; on the other hand, the motion of the nonlinear oscillator jumps from low to high level.
		Note that this phenomenon appears when the nonlinear natural frequency joins the linear one, the system of two coupled oscillators enters an internal resonance \cite{Nayfeh-Mook}.
\end{description}

\subsubsection*{Example 2.}
The second example involves an hysteretic nonlinearity with a rather high level of dissipation (see figures~\ref{fig:Cycle2}) which contrasts with the first example where the damping ratio was quite small.

\begin{figure}[htbp]
    \centering
    \subfigure[Linear oscillator]{\includegraphics[width=.49\textwidth]{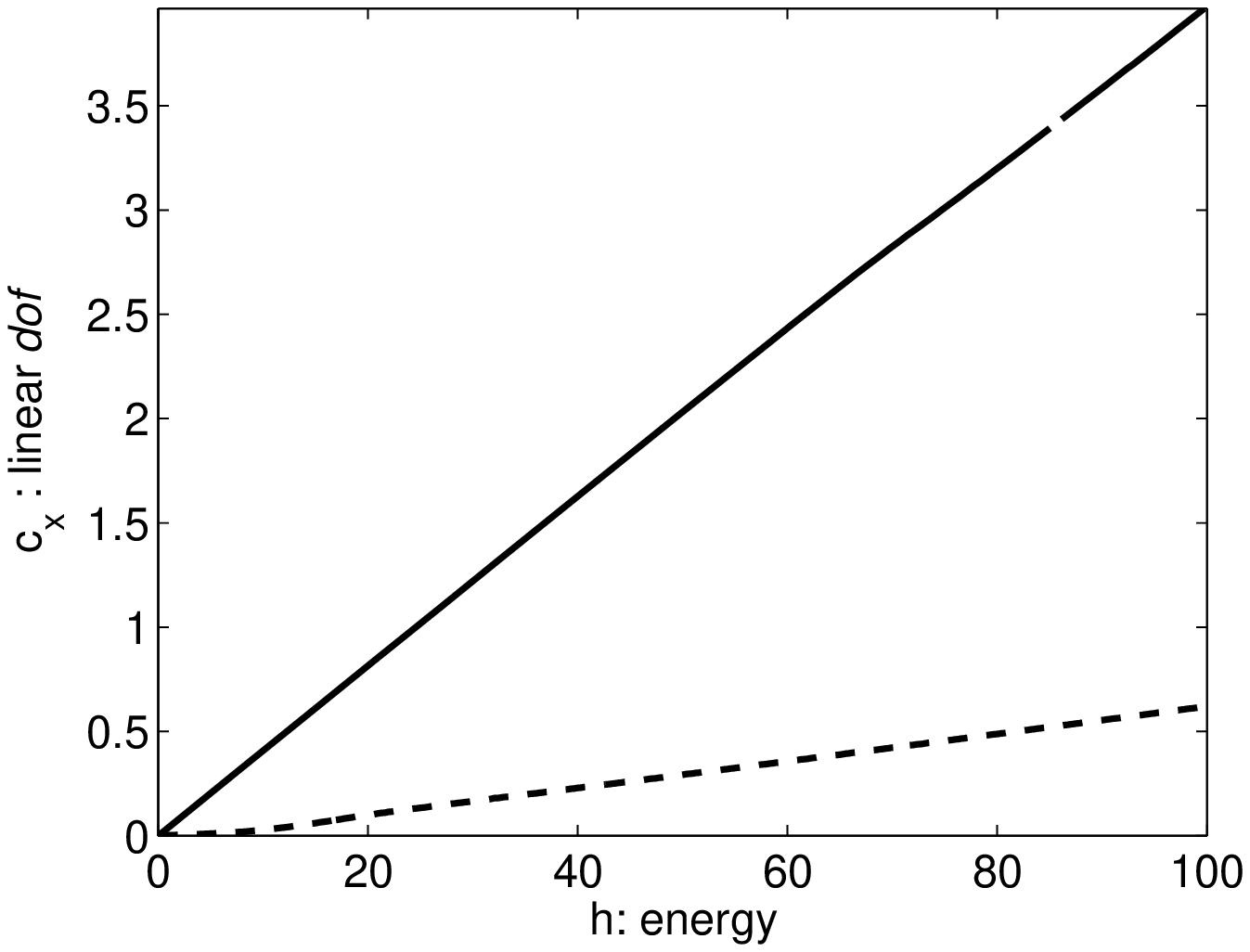}}
    \subfigure[Absorber]{\includegraphics[width=.49\textwidth]{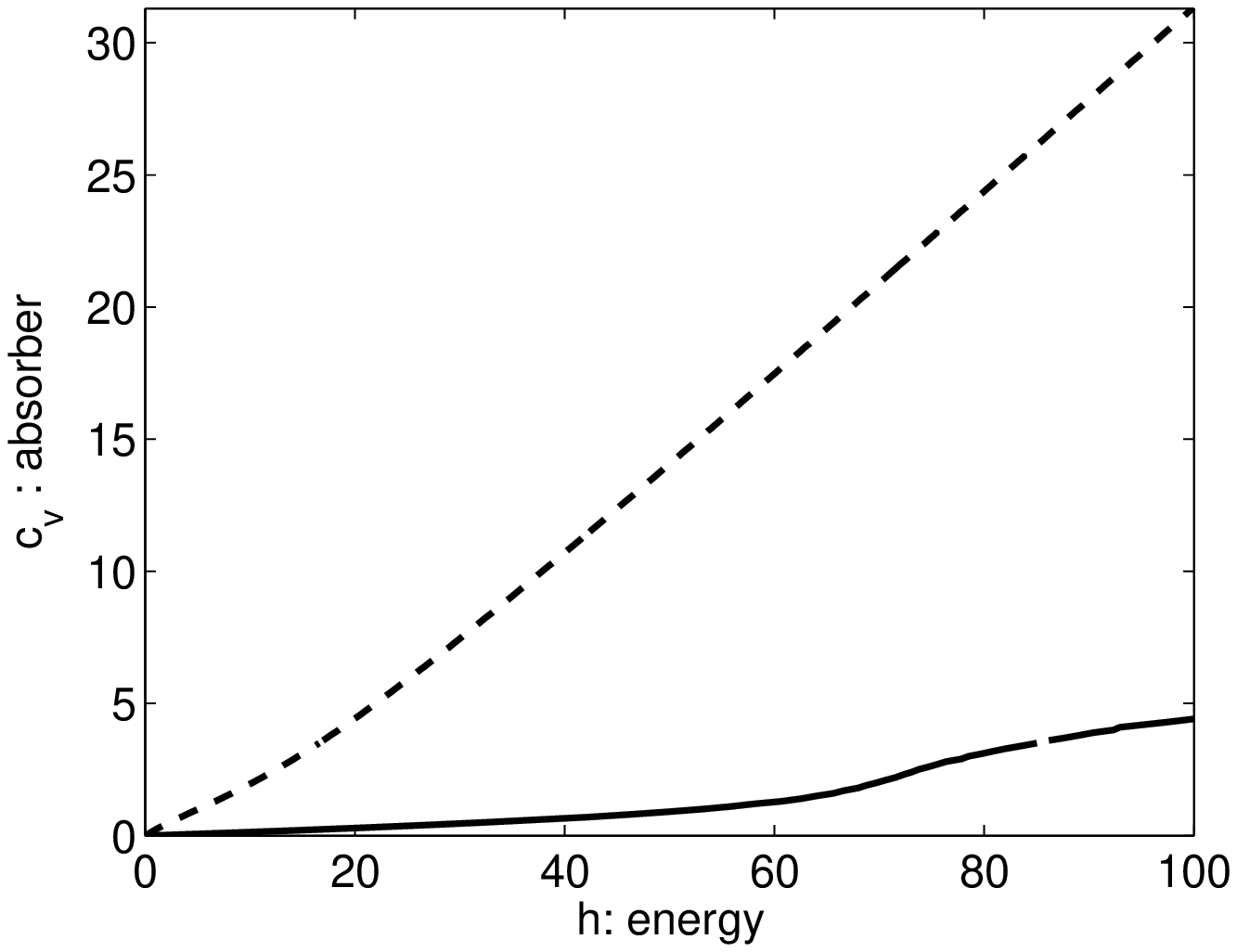}} 
    \subfigure[Eigenfrequency]{\includegraphics[width=.49\textwidth]{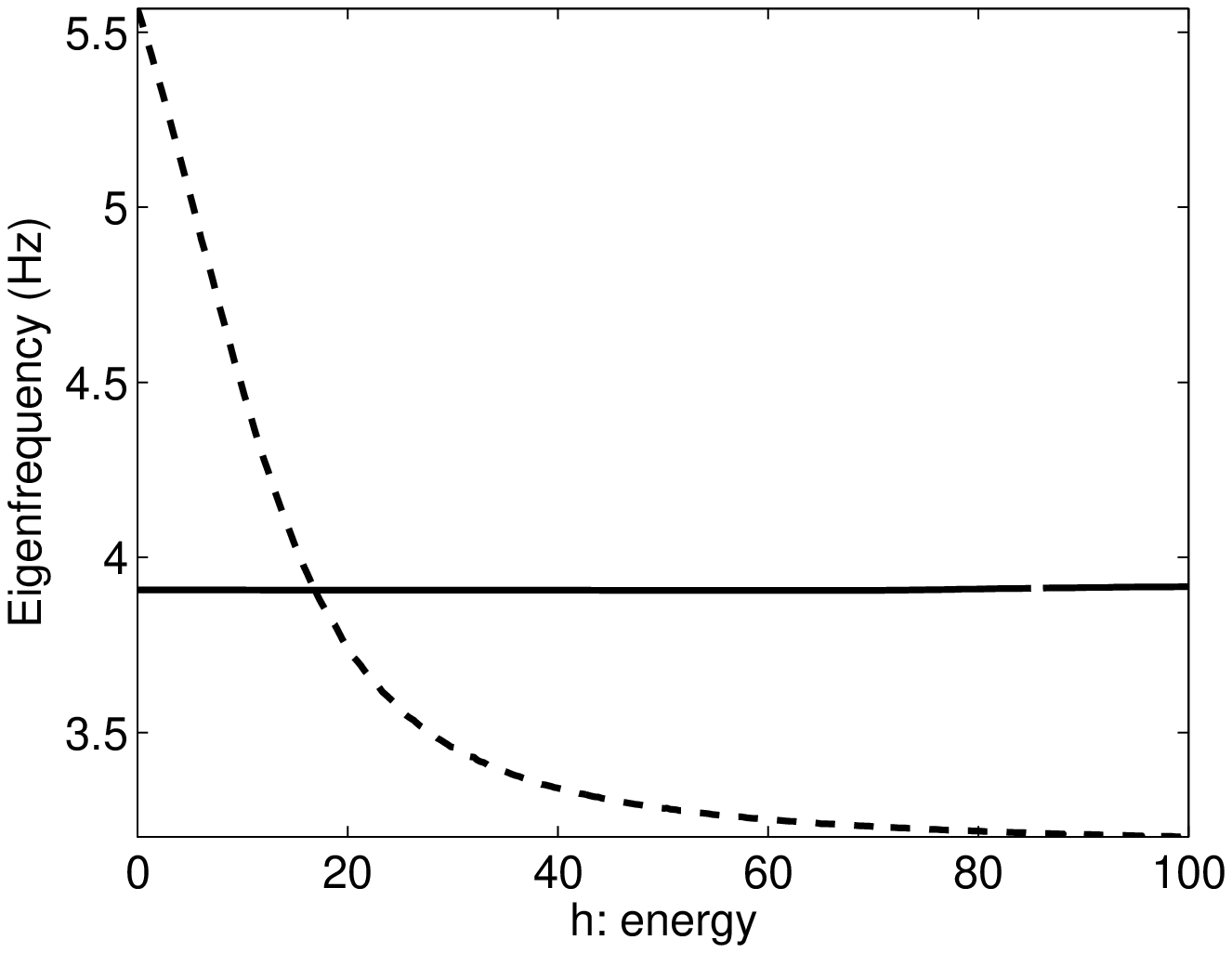}}
    \subfigure[Modal damping]{\includegraphics[width=.49\textwidth]{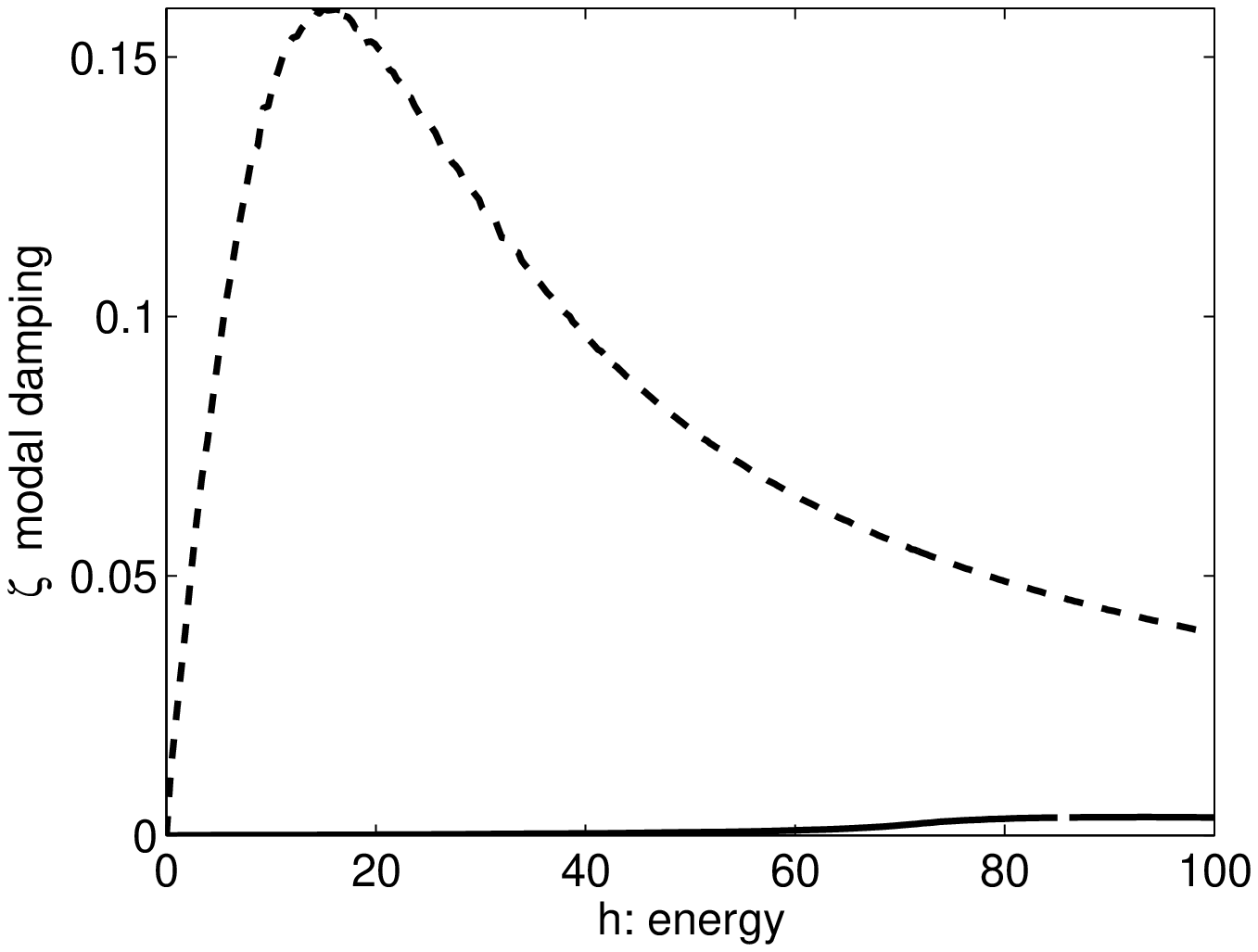}}
    \caption[Nonlinear modes - Example 2]{Example 2: Nonlinear modes}
    \label{fig:ModesNL2}
\end{figure}

The modal quantities are represented in figures~\ref{fig:ModesNL2}.
There are several differences in this example with reference to the previous one.
First note that there is no instability region and no internal resonance.
Then the modal damping curve is quite different from the previous example since the dashed line branch reach higher values of damping especially in the region where the two frequency curves could interact.
In effect, figure~\ref{fig:Cycle2} shows a particular feature of this nonlinearity which is that the level of nonlinearity (in this case, the drop of stiffness) is directly related to the level of damping.
As a consequence, a too high value of nonlinear damping can inhibit the resonance capture and no energy pumping occurs and two oscillators behave quite independently.

This comparison clearly shows this importance of the level of dissipation in the nonlinearity on the efficiency of the energy pumping phenomenon.
However, remember that, in the energy pumping phenomenon, as the vibratory energy gets transferred to the absorber it has to be dissipated.
The optimal absorber would then be the one with the highest damping ratio allowing the resonance capture to take place and leading a fast energy dissipation.
Example 2 displays a case where a high dissipation rate inhibits the energy transfer whereas example 1 shows an optimal design, in term of dissipation, which ensures that both the energy transfer and the energy dissipation are achieved.
\section{Transient free response} \label{sect:Transient}
\begin{figure}[ht]
	\begin{center}
		\subfigure[]{\includegraphics[width=.48\textwidth]{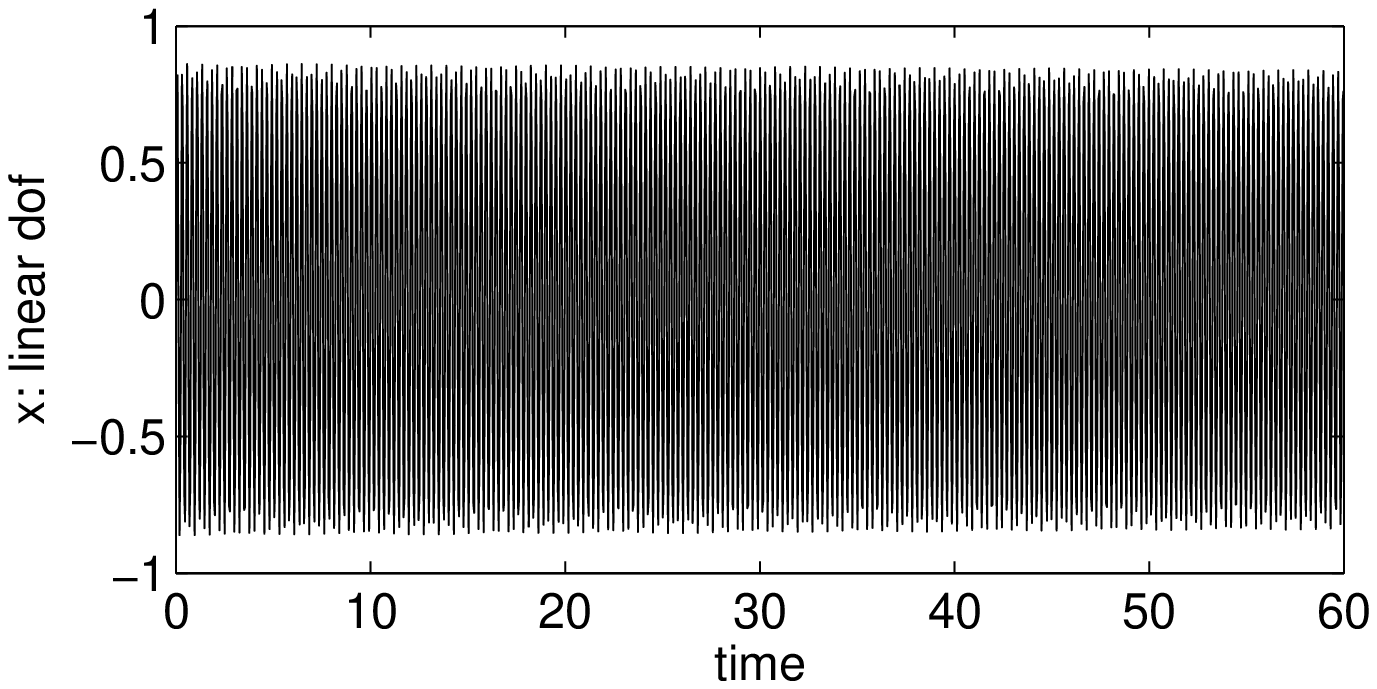}}
		\subfigure[]{\includegraphics[width=.48\textwidth]{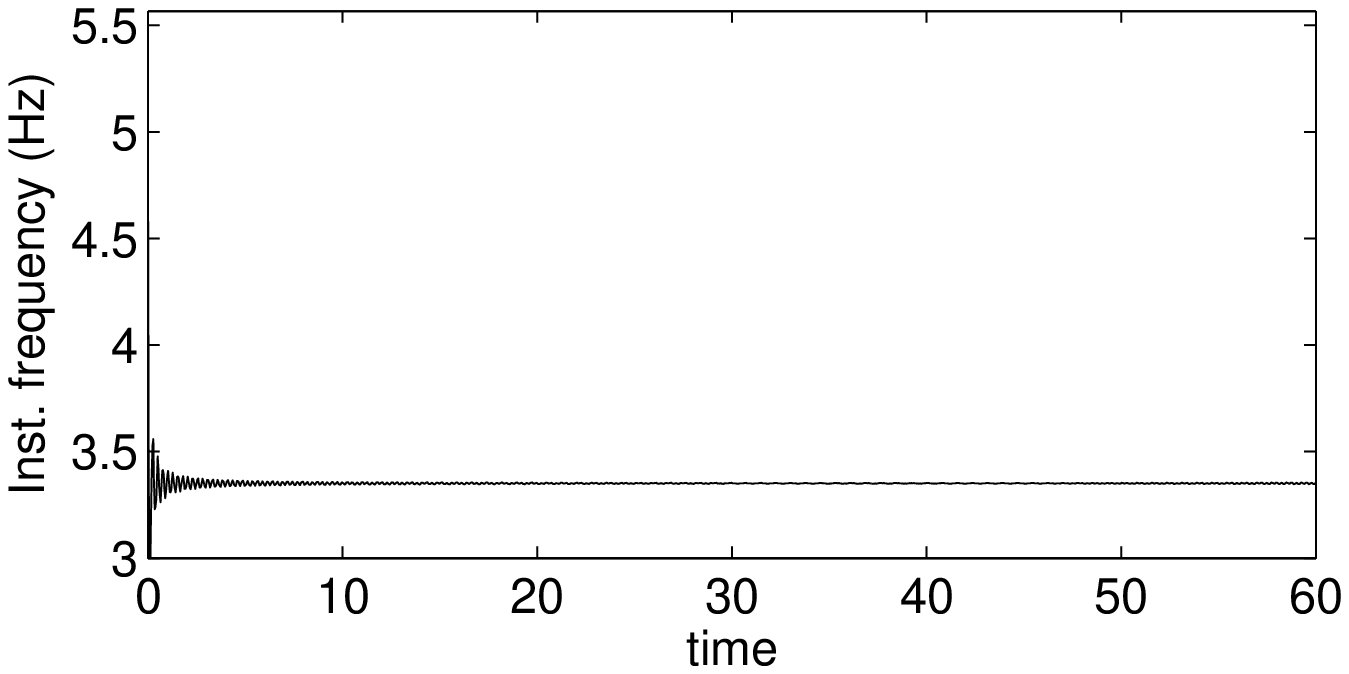}}
	\end{center}
	\caption[Transient response for $\mathcal{H}_0=15$ -- Linear oscillator]{Transient response for $\mathcal{H}_0=15$ -- Linear oscillator; (a)~time history, (b)~instantaneous frequency.}
	\label{fig:TransH15-x}
\end{figure}

\begin{figure}[ht]
	\begin{center}
		\subfigure[]{\includegraphics[width=.48\textwidth]{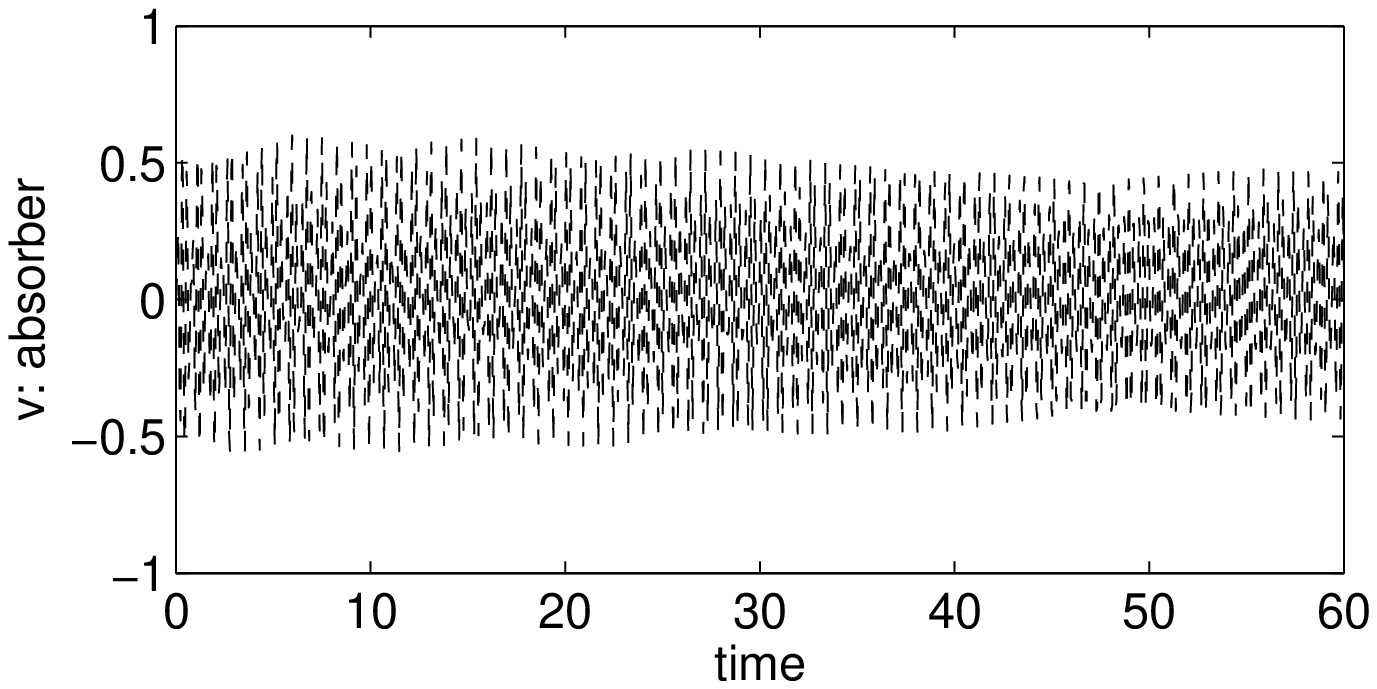}}
		\subfigure[]{\includegraphics[width=.48\textwidth]{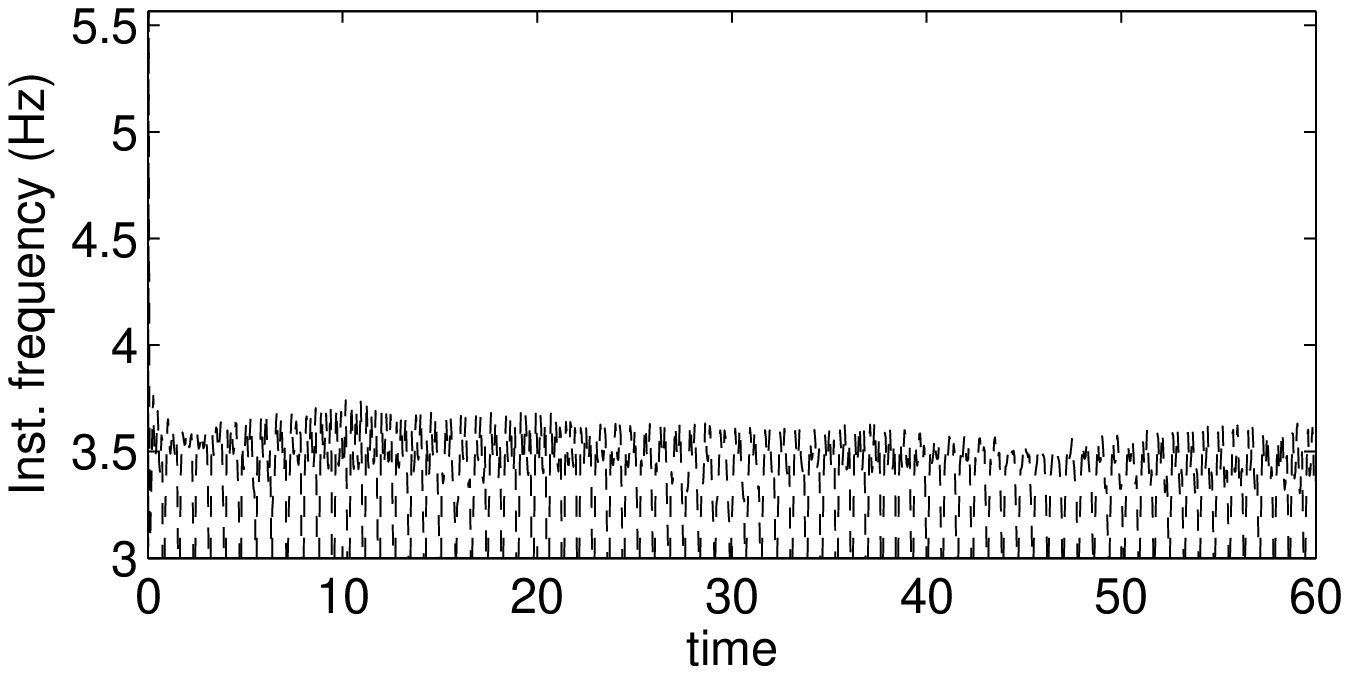}}
	\end{center}
	\caption[Transient response for $\mathcal{H}_0=15$ -- Absorber]{Transient response for $\mathcal{H}_0=15$ -- Absorber; (a)~time history, (b)~instantaneous frequency.}
	\label{fig:TransH15-v}
\end{figure}

\begin{figure}[ht]
	\begin{center}
		\subfigure[]{\includegraphics[width=.48\textwidth]{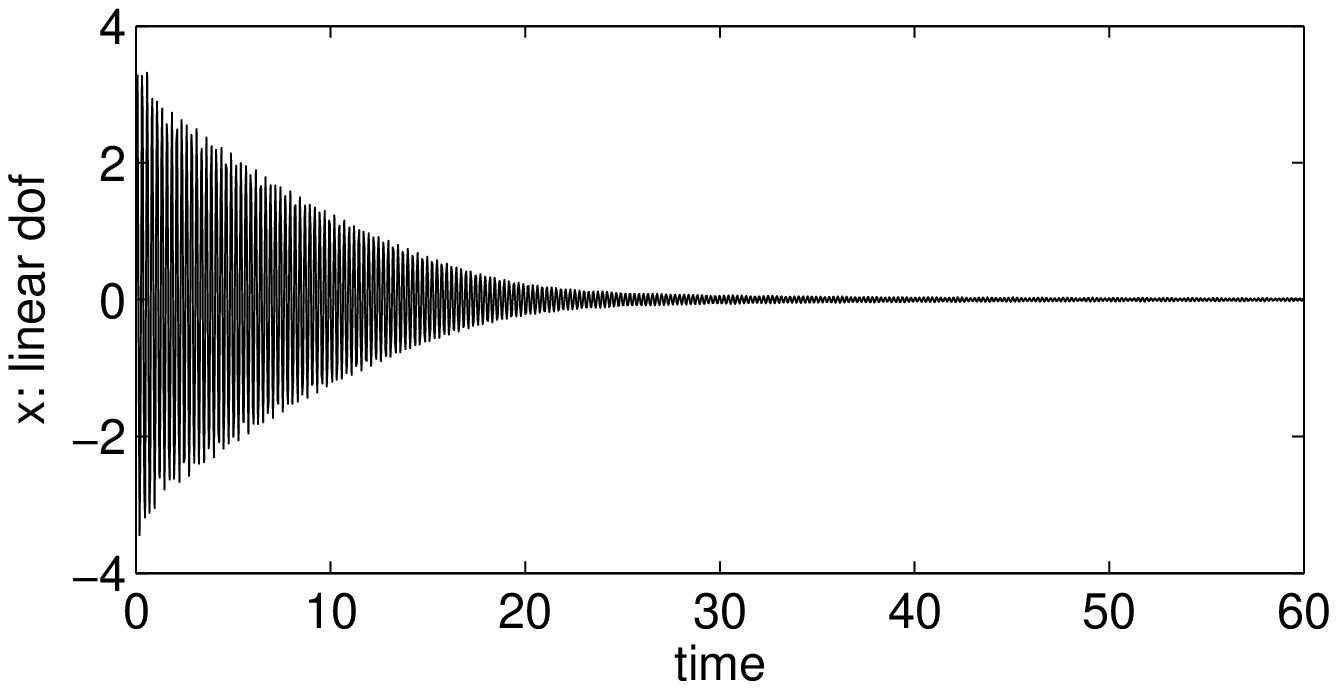}}
		\subfigure[]{\includegraphics[width=.48\textwidth]{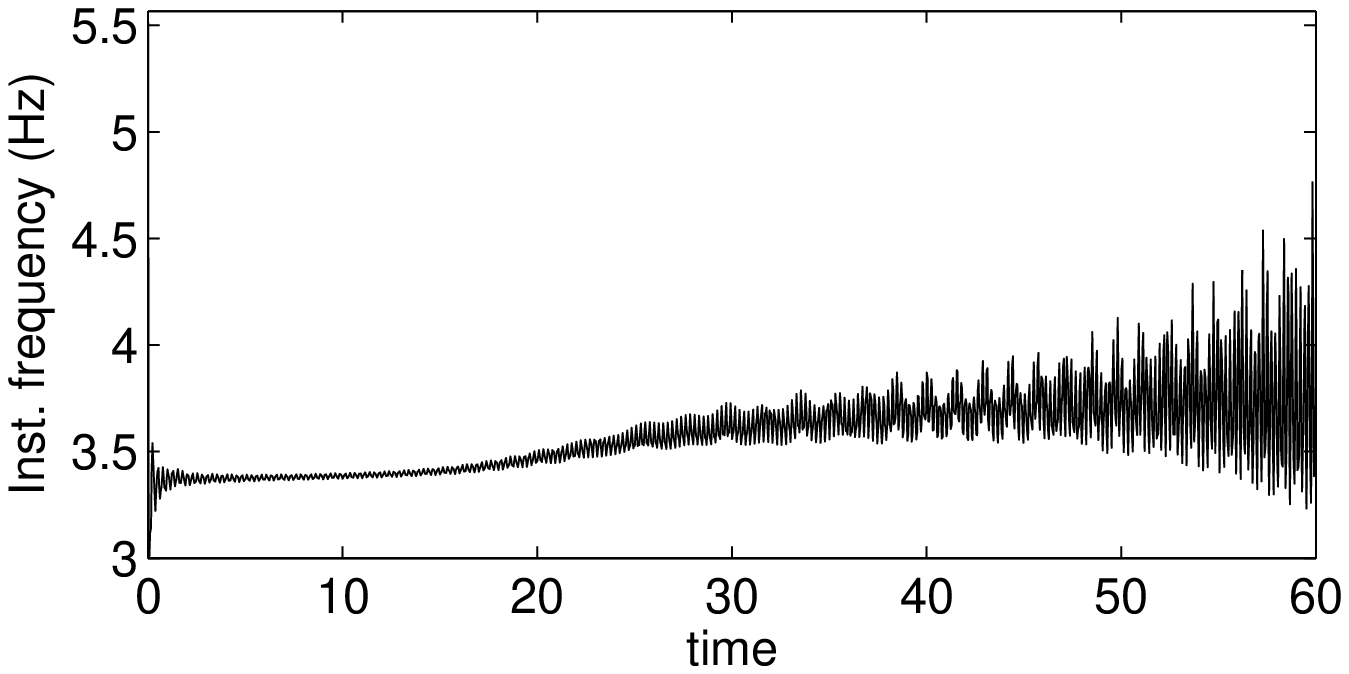}}
	\end{center}
	\caption[Transient response for $\mathcal{H}_0=60$ -- Linear oscillator]{Transient response for $\mathcal{H}_0=60$ -- Linear oscillator; (a)~time history, (b)~instantaneous frequency.}
	\label{fig:TransH60-x}
\end{figure}

\begin{figure}[ht]
	\begin{center}
		\subfigure[]{\includegraphics[width=.48\textwidth]{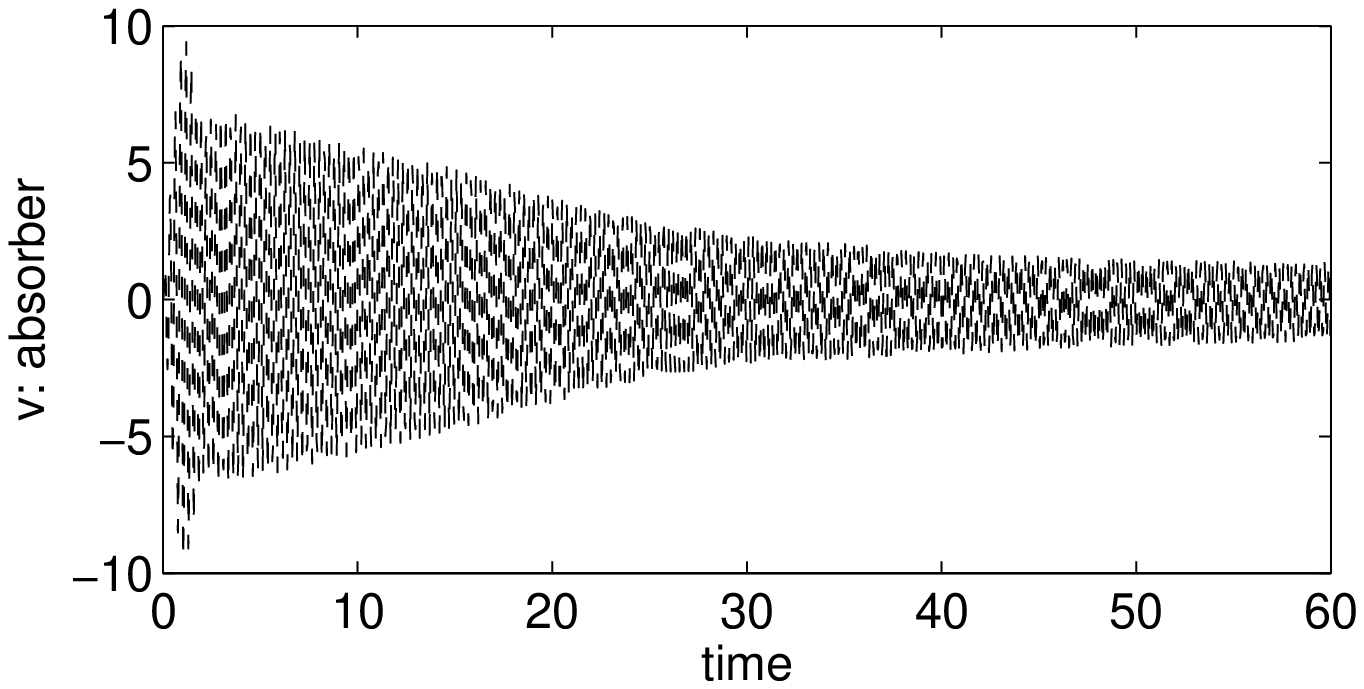}}
		\subfigure[]{\includegraphics[width=.48\textwidth]{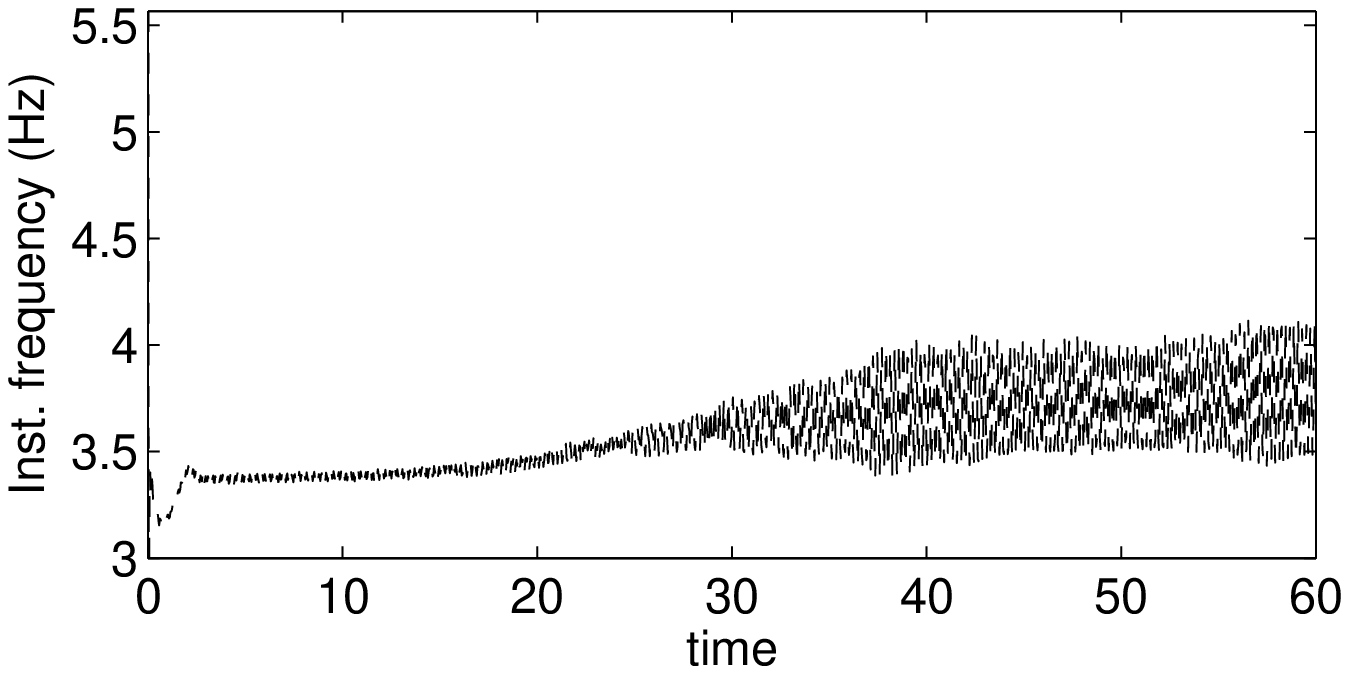}}
	\end{center}
	\caption[Transient response for $\mathcal{H}_0=60$ -- Absorber]{Transient response for $\mathcal{H}_0=60$ -- Absorber; (a)~time history, (b)~instantaneous frequency.}
	\label{fig:TransH60-v}
\end{figure}

In this section, we provide examples of transient free responses which highlight the energy pumping phenomenon and underline the efficiency of the previous modal analysis predictions.
We focus on the first example nonlinearity ($\gamma=10^{-3}$) which appears to be more efficient according to the non-linear modes predictions.
The exact system of equations~(\ref{eq:mvtR}) (with $f(t)=0$) was used with the Bouc-Wen restoring force defined by equation~(\ref{eq:BoucWen}); we impose the initial following initial conditions:
\begin{equation}
	\dot x(0) = \sqrt{\dfrac{2 \mathcal{H}_0}{M}}, \ x(0)=0, \ \dot v (0) = 0 \ \mbox{ and } v (0) = 0
\end{equation}
where $\mathcal{H}_0$ is the initial energy of the system.
These initial conditions simulate an impact on the main mass. 
The results of two simulations with different initial energy input are depicted in figures~\ref{fig:TransH15-x} to \ref{fig:TransH60-v}.
Beside from the displacements $x$ and $v$ history, the history of the instantaneous frequency is also plotted for both oscillators.
The instantaneous frequency was calculated using an Hilbert transform \cite{Worden-Tomlinson}.



For $\mathcal{H}_0 = 15$ (figures~\ref{fig:TransH15-x} and \ref{fig:TransH15-v}), the input of energy is smaller than the bifurcation characteristic level. 
Hence, as predicted by the nonlinear modes (see figure \ref{fig:ModesNL1}) and in accordance with the initial conditions, only the solid line branch of the modes is realizable. 
This means that the vibratory energy remains in the main oscillator: no energy pumping occurs.

For $\mathcal{H}_0 = 60$ (figures~\ref{fig:TransH60-x} and \ref{fig:TransH60-v}), the initial energy is greater than the critical bifurcation level. 
As a consequence, the dashed line branch in figures \ref{fig:ModesNL1} is feasible. 
The energy is rapidly transferred from the main mass to the absorber. 
To confirm the predictions of the nonlinear modes, we can watch the history of the instantaneous frequency: in the early moments (after a short transient period), the instantaneous frequencies from the linear and nonlinear oscillator join, then (as the global energy decreases) both of them increase. 
This is in accordance with the evolution of the nonlinear natural frequencies of the modes.

\section{Steady-State Forced Response} \label{sect:ForcedResp}
This final part is dedicated to forced resonance phenomena.
We were, in particular, interested in periodically forced regimes.
In contrast with transient phenomena, in steady-state forced vibration, failures generally occur because of high cycle fatigue effects.
Therefore, control of forced vibrations is of primary importance in many fields of mechanical engineering.
It will be demonstrated that the energy pumping phenomenon can achieve this control function.

\subsection{Derivation of periodic solutions}
We consider an harmonic excitation, $f(t)=P\cos\omega t$ in the system~(\ref{eq:mvtR}).
The derivation of periodic solution uses the averaging procedure of section \ref{sect:AsymptoticAnalysis}, in which the two displacements variables take the form of (\ref{eq:solapprox}), {\it ie}:
\begin{equation} 
	x(\tau,\eta)=a_x(\eta) \cos(\tau +\varphi_x(\eta)) \mbox{ and } v(\tau,\eta)=a_v(\eta) \cos(\tau +\varphi_v(\eta))
	\label{eq:SS-solapprox}
\end{equation}
where $\tau=\omega t$.

The term $G(a,\varphi,\tau)$ in equations~(\ref{eq:SlowStand}) and (\ref{eq:SlowMoy}) is respectively for $x$ and $v$:
\begin{subequations}
   \begin{equation}
      G_x(a_x,\varphi_x,\tau)=-\lambda_0\omega a_x\sin(\tau +\varphi_x) +  \omega_0^2a_x\cos(\tau +\varphi_x) - \epsilon_0 a_v\cos(\tau +\varphi_v) - P\cos \tau
   \end{equation}
   \begin{equation}
      G_v(a_v,\varphi_v,\tau)=-\lambda_1\omega a_v\sin(\tau +\varphi_v) +  \omega_1^2a_v\cos(\tau +\varphi_v) - \epsilon_1 a_x\sin(\tau +\varphi_x) + a_r(a_v,\varphi_v) \cos(\tau+\varphi_r(a_v,\varphi_v))
   \end{equation}
\end{subequations}

Then applying the relations~(\ref{eq:SlowMoy}) to the present example, one obtains:
\begin{subequations} \label{eq:SlowMoyApp}
   \begin{equation}
      a_x^{\prime} = -\dfrac{\epsilon_0}{2\omega}a_v\sin(\varphi_x-\varphi_v) -\frac{\lambda_0}{2}a_x -\frac{P}{2\omega}\sin\varphi_x
   \end{equation}
   \begin{equation}
      a_v^{\prime} = \dfrac{\epsilon_1}{2\omega}a_x\sin(\varphi_x-\varphi_v) -\frac{\lambda_1}{2}a_v +\dfrac{1}{2\omega} a_r\sin(\varphi_v-\varphi_r)
   \end{equation}
   \begin{equation}
      a_x \varphi_x^{\prime} = \frac{\omega_0^2-\omega^2}{2\omega} -\dfrac{\epsilon_0}{2\omega}a_v\cos(\varphi_x-\varphi_v)-\frac{P}{2\omega}\cos\varphi_x
   \end{equation}
   \begin{equation}
      a_v \varphi_v^{\prime} = \frac{\omega_1^2-\omega^2}{2\omega} -\dfrac{\epsilon_1}{2\omega}a_x\cos(\varphi_x-\varphi_v) +\dfrac{1}{2\omega} a_r\cos(\varphi_v-\varphi_r)
   \end{equation}
\end{subequations}
As for the free response, the fixed points of system~(\ref{eq:SlowMoyApp}) are to the main approximation to the forced response. Next, these forced response will be investigated for several values of load amplitude $P$.

\subsection{Numerical results}
The following results were obtained by numerically solving the nonlinear system~(\ref{eq:SlowMoyApp}) using Newton-like solver combine with an arc-length continuation method \cite{Nayfeh-Applied}. 
We used the first example Bouc-Wen hysteretic cycle ($\gamma = 10^{-3}$, figure~\ref{fig:Cycle2}).

The nonlinear frequency response are compared with the frequency responses of the linear system alone; the backbone curves have also been represented in order to see how the prediction of the nonlinear modes are in accordance with the forced response.

\begin{figure}[htbp]
	\centering
	\subfigure[Linear oscillator]{\includegraphics[width=.49\textwidth]{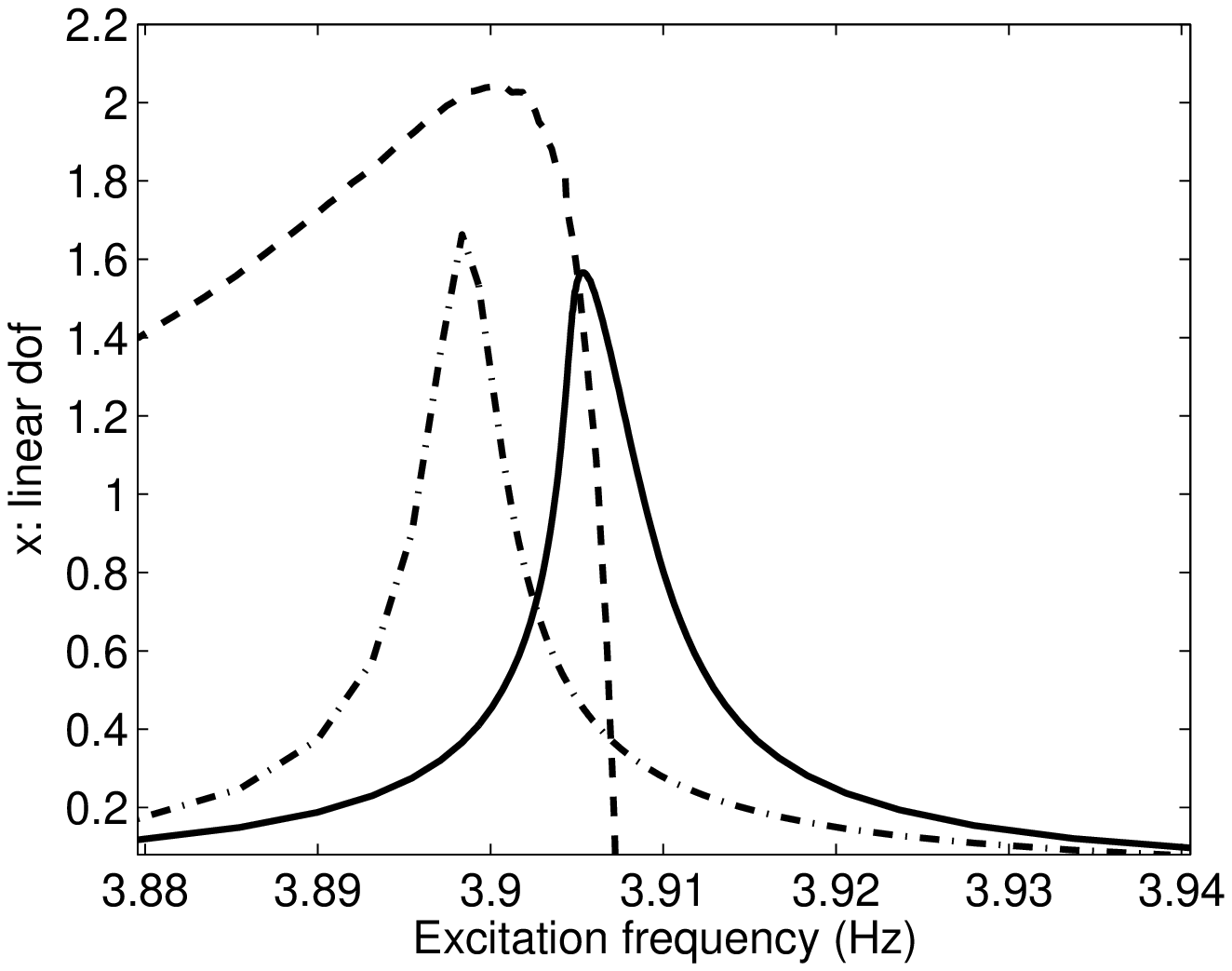}}
	\subfigure[Absorber]{\includegraphics[width=.49\textwidth]{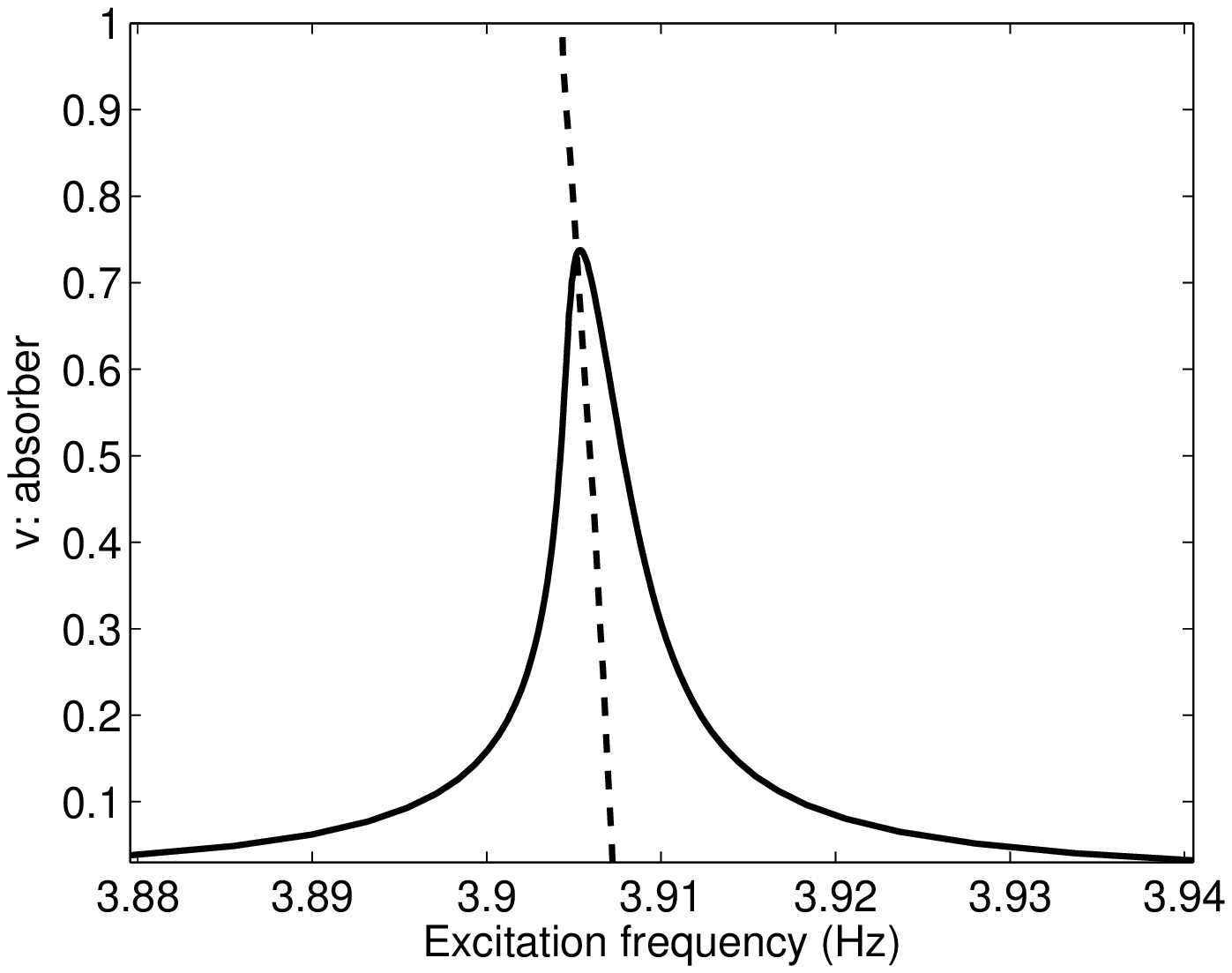}}
	\caption[Forced response -- Low energy]{Forced response -- Low energy: (------), nonlinear response, ($-\cdot-\cdot$), linear response (no absorber), ($- - -$), backbone curves}
	\label{fig:RF_LowEnergy}
\end{figure}

In the first example (figure~\ref{fig:RF_LowEnergy}), the forcing level is quite small and we can see in both the linear and nonlinear oscillator have a linearizable behaviour. 

\begin{figure}[htbp]
	\centering
	\subfigure[Linear oscillator]{\includegraphics[width=.48\textwidth]{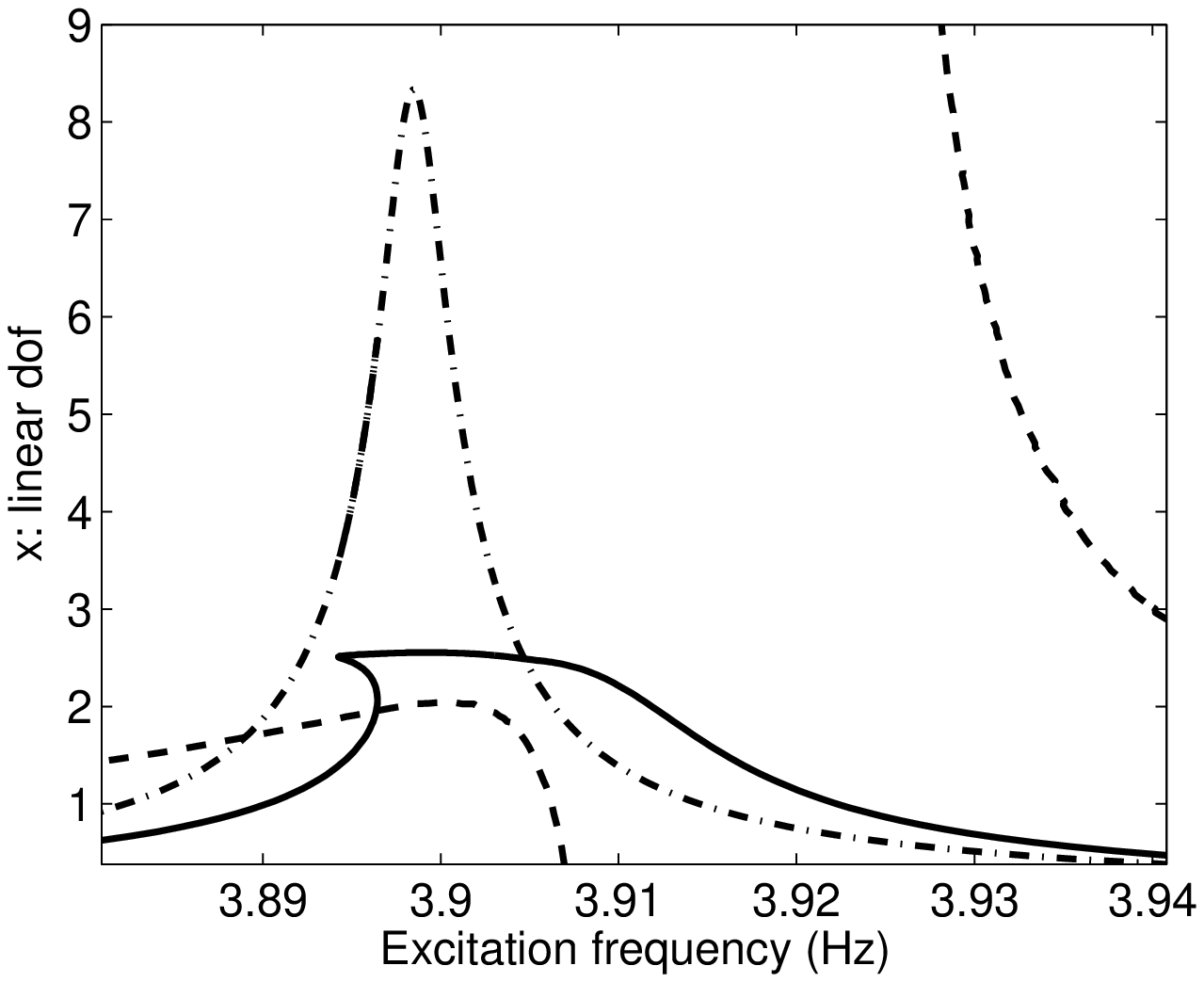}}
	\subfigure[Absorber]{\includegraphics[width=.49\textwidth]{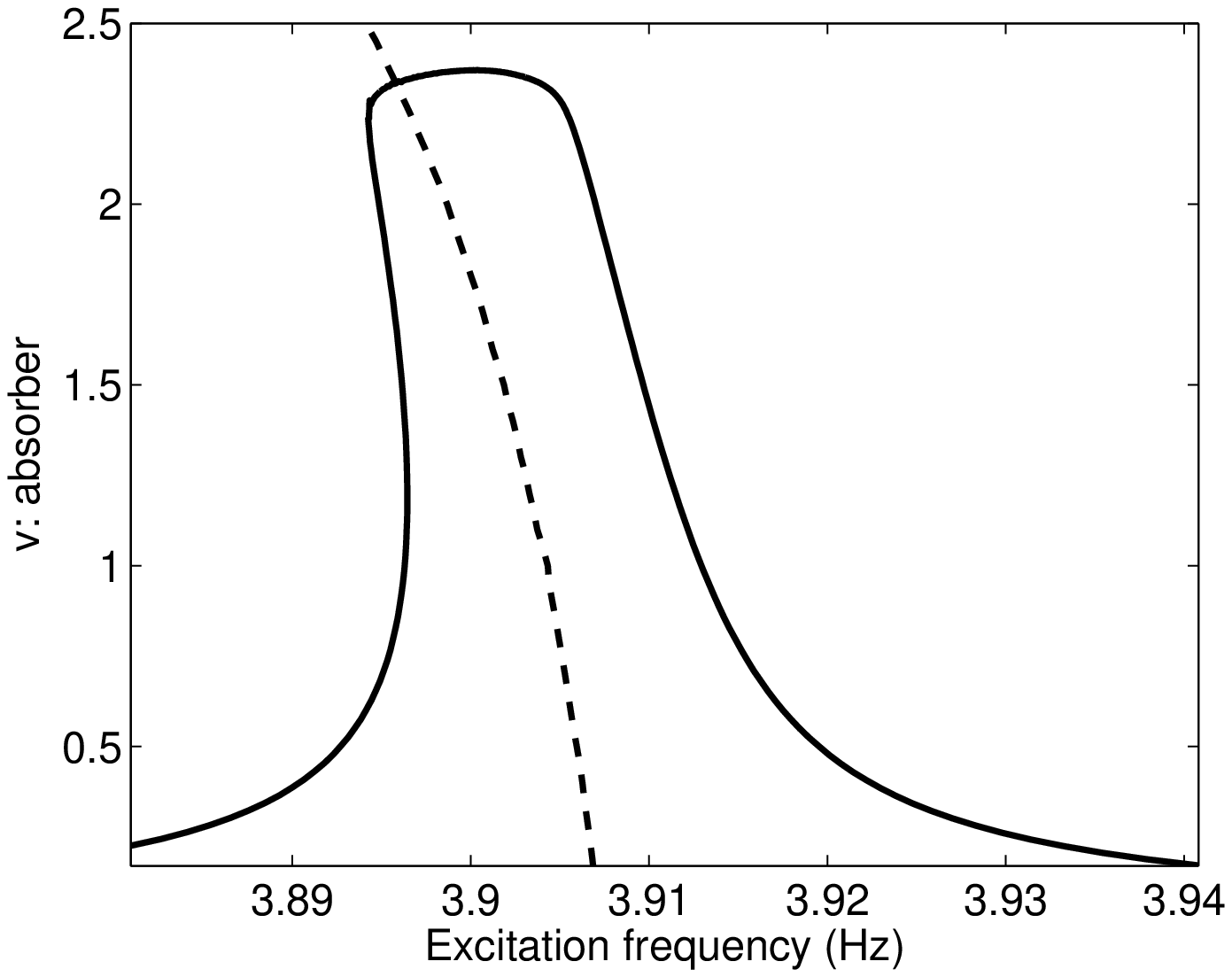}}
	\caption[Forced response -- High energy]{Forced response -- High energy: (------),  nonlinear response, ($-\cdot-\cdot$), linear response (no absorber), ($- - -$), backbone curves}
	\label{fig:RF_HighEnergy}
\end{figure}

As the level of the excitation is increased, the system's behaviour differs from the linear case and several interesting phenomena appear. An example of nonlinear response is plotted in figure~\ref{fig:RF_HighEnergy}. 
In this case, the vibratory energy is strongly localized in the nonlinear oscillator (absorber) in the vicinity of the resonance peak. 
The absorber appears to be efficient. 
One can also notices that the nonlinear response remains quite close to the backbone curves which attests the quality of the prediction of the nonlinear modes.

\section{Conclusions}
The results of an investigation on the dynamics of a small nonlinear oscillator weakly coupled with a linear oscillator were presented.
This investigation focused on hysteretic nonlinearity using a Bouc-Wen differential model.
It was shown that the absorber can act as an energy sink when it is properly designed; in particular it appears that the level of nonlinearity and the level of damping are important factor for the efficiency of the device.
In order to derive approximate solutions to the nonlinear problem, an averaging strategy was used. 
Investigations of the free and forced responses are presented and, in both cases, it appears that when the energy of the system is sufficient some localization of the vibratory energy in the nonlinear absorber appears.
With the examples presented in this paper, we have seen that the nonlinear modes were representative of the behaviour of nonlinear dissipative system in free response as in forced response.


\end{document}